\documentclass[onecolumn]{aastex60}
\pdfoutput=1 
\usepackage{amsmath,amstext}
\usepackage[T1]{fontenc}
\usepackage{apjfonts} 
\usepackage[figure,figure*]{hypcap}
\usepackage[utf8]{inputenc}
\usepackage{array}
\usepackage{makecell}
\usepackage{mathrsfs}
\usepackage{mathtools}
\usepackage{lineno}
\usepackage{footmisc}

\shorttitle{A Search for Neutrinos from Fast Radio Bursts}
\shortauthors{M. Aartsen et al.}

\begin{document}
\email{analysis@icecube.wisc.edu}
\title{A Search for Neutrino Emission from Fast Radio Bursts with Six Years of IceCube Data}


\author{
IceCube Collaboration\footnote[1]{\email{analysis@icecube.wisc.edu}}:
M.~G.~Aartsen\altaffilmark{1},
M.~Ackermann\altaffilmark{2},
J.~Adams\altaffilmark{1},
J.~A.~Aguilar\altaffilmark{3},
M.~Ahlers\altaffilmark{4},
M.~Ahrens\altaffilmark{5},
I.~Al~Samarai\altaffilmark{6},
D.~Altmann\altaffilmark{7},
K.~Andeen\altaffilmark{8},
T.~Anderson\altaffilmark{9}
% ...
}
\altaffiltext{1}{Dept.~of Physics and Astronomy, University of Canterbury, Private Bag 4800, Christchurch, New Zealand}
\altaffiltext{2}{DESY, D-15738 Zeuthen, Germany}
\altaffiltext{3}{Universit\'e Libre de Bruxelles, Science Faculty CP230, B-1050 Brussels, Belgium}
\altaffiltext{4}{Niels Bohr Institute, University of Copenhagen, DK-2100 Copenhagen, Denmark}
\altaffiltext{5}{Oskar Klein Centre and Dept.~of Physics, Stockholm University, SE-10691 Stockholm, Sweden}
\altaffiltext{6}{D\'epartement de physique nucl\'eaire et corpusculaire, Universit\'e de Gen\`eve, CH-1211 Gen\`eve, Switzerland}
\altaffiltext{7}{Erlangen Centre for Astroparticle Physics, Friedrich-Alexander-Universit\"at Erlangen-N\"urnberg, D-91058 Erlangen, Germany}
\altaffiltext{8}{Department of Physics, Marquette University, Milwaukee, WI, 53201, USA}
\altaffiltext{9}{Dept.~of Physics, Pennsylvania State University, University Park, PA 16802, USA}
% ...

\begin{abstract}
We present a search for coincidence between IceCube TeV neutrinos and fast radio bursts (FRBs). During the search period from 2010 May 31 to 2016 May 12, a total of 29 FRBs with 13 unique locations have been detected in the whole sky. An unbinned maximum likelihood method was used to search for spatial and temporal coincidence between neutrinos and FRBs in expanding time windows, in both the northern and southern hemispheres. No significant correlation was found in six years of IceCube data. Therefore, we set upper limits on neutrino fluence emitted by FRBs as a function of time window duration. We set the most stringent limit obtained to date on neutrino fluence from FRBs with an $E^{-2}$ energy spectrum assumed, which is 0.0021 GeV cm$^{-2}$ per burst for emission timescales up to \textasciitilde10$^2$ seconds from the northern hemisphere stacking search.
\end{abstract}


\section{Introduction} \label{intro}

Fast radio bursts (FRBs) are a new class of astrophysical phenomenon characterized by bright broadband radio emission lasting only a few milliseconds. Since the first FRB discovered in 2007 in archival data from the Parkes Radio Telescope \citep{Lorimer777}, more than 20 FRBs have been detected by a total of five observatories \citep{Spitler:2014fla, Masui:2015kmb, Caleb:2017vbk, Bannister:2017sie}. This rules out the hypothesis of instrumental or  terrestrial origin of these phenomena. The number of FRBs detected together with the duration and solid angle searched imply an all-sky FRB occurrence rate of a few thousand per day \citep{Thornton:2013iua, Spitler:2014fla}, which is consistent with 10\% of the supernova rate \citep{Murase:2016sqo}. The burst durations suggest that FRB progenitors are very compact, with light-transit distances on the order of hundreds of kilometers. The dispersion measures -- the time delay of lower frequency signal components, which is proportional to the column density of free electrons along the line of sight -- of the detected FRBs are significantly greater than the Milky Way alone could provide~\citep{Cordes:2016rpr}, and the majority of sources have been detected at high Galactic latitudes, indicating extragalactic origin. The distances of the FRBs extracted from their dispersion measures, however, are only upper limits and precise measurements are yet to be determined, most likely from multi-wavelength observations. 

The nature of FRBs is still under heated debate, and a multitude of models have been proposed for the FRB progenitors, the majority of which involve strong magnetic fields and leptonic acceleration. Some models predict millisecond radio bursts from cataclysmic events such as dying stars \citep{Falcke:2013xpa}, neutron star mergers \citep{Totani:2013lia}, or evaporating
black holes \citep{rees1977better}. In 2015, 16 additional bursts were detected from the direction of FRB 121102 \citep{Spitler:2014fla, Scholz:2016rpt}, spaced out non-periodically by timescales ranging from seconds to days. This indicates that the cataclysmic scenario is not true at least for this repeating FRB. A multi-wavelength follow-up campaign identified this FRB’s host dwarf galaxy at a distance of \textasciitilde1 Gpc \citep{Chatterjee:2017dqg}. It is unclear whether FRB 121102 is representative of FRBs as a source class or if repetitions are possible for only a subclass of FRBs.

While leptonic acceleration is typically the default assumption for FRB emission in most models, hadronic acceleration is also possible in the associated regions of the progenitors, which would lead to production of high-energy cosmic rays and neutrinos \citep{Li2013}. It has been proposed that cosmological FRBs could link to exotic phenomena such as oscillations of superconducting cosmic strings \citep{Ye:2017lqn}, and some authors predict that such cosmic strings could also produce ultra-high energy cosmic rays and neutrinos, from super heavy particle decays \citep{Berezinsky:2009xf, Lunardini:2012ct}. Therefore, both multi-wavelength and multi-messenger follow-ups can provide crucial information to help decipher the origin of FRBs. Here, the IceCube telescope offers the opportunity to search for neutrinos correlated with FRBs.


The IceCube Neutrino Observatory consists of 5160 digital optical modules (DOMs) instrumenting one cubic kilometer of Antarctic ice from depths of 1450 m to 2450 m at the geographic South Pole \citep{Aartsen:2016nxy}. Charged products of neutrino interactions in the ice create Cherenkov photons which are observed by the DOMs and allow the reconstruction of the initial neutrino energy, direction, and interaction type. Charged-current muon neutrino interactions create muons, which travel along straight paths in the ice, resulting in events with directional resolution $\lesssim 1^{\circ}$ at energies above 1 TeV \citep{maunu_phd}. The detector – fully installed since 2010 – collects data from the whole sky with an up-time higher than 99\% per year, enabling real-time alerts to other instruments and analysis of archival data as a follow-up to interesting signals detected by other observatories.

IceCube has discovered a diffuse astrophysical neutrino flux in the TeV to PeV energy range \citep{IceCubePeV, HESE1, HESE2, mese, muonNeutrinos, numu6year}. The arrival directions of these neutrinos are consistent with an isotropic distribution, indicating a majority of them have originated from extragalactic sources. Although tau neutrinos are yet to be identified among the observed flux \citep{Aartsen:2015dlt}, the flavor ratio is found to be consistent with $\nu_{e}$ : $\nu_{\mu}$ : $\nu_{\tau}$ = 1 : 1 : 1 from analyses which combined multiple data sets \citep{globalFit} and with events starting inside the detector for all flavor channels. \citep{PhysRevLett.114.171102, Aartsen:2017mau}. Close-to-equal flavor ratio is another feature of astrophysical neutrinos which have traversed astronomical distances and hence have reached full mixing \citep{PhysRevLett.115.161303, PhysRevLett.115.161302}. While the astrophysical neutrino flux has been detected in multiple channels with high significance, neither clustering in space or time nor cross correlations to catalogs have been found \citep{Aartsen:2016oji}. The once promising sources for high-energy neutrinos such as gamma ray bursts \citep{Abbasi:2012zw, aartsen2015search, Aartsen:2017wea} and blazars \citep{Aartsen:2016lir} have been disfavored as the major contributors to the observed flux. To date, the origin of the astrophysical neutrinos remains a mystery.

In \cite{Fahey:2016czk}, an analysis of four FRBs with one year of IceCube data was reported. Here we present the results of a more sophisticated study in search of high-energy neutrinos from 29 FRBs using the IceCube Neutrino Observatory. The paper is structured as follows: Section~\ref{event_sample} describes the event sample used. We then discuss the analysis method, search strategies and background modeling in Section~\ref{ana_method}. In Section~\ref{sensitivity}, we present the sensitivities and discovery potentials based on the analysis method and search strategies established in Section~\ref{ana_method}. We then report the final results and their interpretation in Section~\ref{results}. Finally, we conclude and discuss the future prospects for FRB follow-ups with IceCube in Section~\ref{conclusion}.

\section{Event Sample} \label{event_sample}

The data used in this analysis are assembled from muon neutrino candidate events selected in previous analyses in search of prompt neutrino coincidence with gamma ray bursts (GRBs) \citep{aartsen2015search, Aartsen:2017wea}. It consists of ten data sets: five years of data from the northern hemisphere and five from the southern hemisphere (Table~\ref{tab:data}). Due to the effects of atmospheric muon contamination, which are strong in the south and negligible in the north, the data samples are constructed in two ``hemispheres'' separated at a declination of $\delta=-5^\circ$.  The northern selection extends to $-5^\circ$ rather than $0^\circ$ declination because there is still sufficient Earth overburden at $-5^\circ$ for efficient absorption of atmospheric muons.

\begin{table}[t]
\caption{For IceCube data during which an FRB was detected, the event rates, numbers of events, and respective livetimes are shown. Here, "IC79" indicates the first year of data used in this analysis, when the IceCube array consisted of 79 strings; "IC86-1", "IC86-2", etc. denote subsequent years of data from the completed 86-string array. The median angular uncertainty among events in each sample is given as a 90\% containment radius, assuming each event reconstruction to have a 2-D Gaussian point-spread function. Since the event reconstruction becomes more accurate for higher energy events, the southern data sets have smaller median angular uncertainties as a consequence of harder energy cuts to reduce atmospheric background. Year-to-year variations in event rate and $\sigma_{90\%}$ are the result of event selection methods aimed to maximize sensitivity independently for each data set's corresponding set of sources in a previous search for GRBs, as described in Section~\ref{event_sample}.}
\vspace{.1in}
\centering
\begin{tabular}{ c c c c c c c }
\hline
\hline
Northern ($\delta>-5^\circ$) Data & Start date & End date & Rate (mHz) & Events & Livetime (days) & $\sigma_{90\%}$ \\
\hline
\hline
IC86-1 & 2011-05-13 & 2012-05-15 & 3.65 & 107,612 & 341.9 & 2.13$^\circ$ \\
\hline 
IC86-2 & 2012-05-15 & 2013-05-02 & 5.50 & 157,754 & 332.2 & 2.68$^\circ$\\
\hline 
IC86-3 & 2013-05-02 & 2014-05-06 & 6.20 & 193,320 & 362.2 & 2.79$^\circ$\\
\hline 
IC86-4 & 2014-05-06 & 2015-05-15 & 6.17 & 197,311 & 369.8 & 2.79$^\circ$\\
\hline 
IC86-5 & 2015-05-15 & 2016-05-12 & 6.07 & 186,600 & 356.8 & 2.83$^\circ$\\
\hline 
\hline
Southern ($\delta<-5^\circ$) Data & Start date & End date & Rate (mHz) & Events & Livetime (days) & $\sigma_{90\%}$ \\
\hline
\hline
IC79 & 2010-05-31 & 2011-05-13 & 2.46 & 67,474 & 314.6 & 1.02$^\circ$ \\
\hline 
IC86-1 & 2011-05-13 & 2012-05-15 & 1.90 & 58,982 & 359.6 & 1.10$^\circ$\\
\hline 
IC86-2 & 2012-05-15 & 2013-05-02 & 3.18 & 91,485 & 328.6 & 1.05$^\circ$\\
\hline 
IC86-3 & 2013-05-02 & 2014-05-06 & 3.23 & 100,820 & 358.6 & 1.04$^\circ$\\
\hline 
IC86-4 & 2014-05-06 & 2015-05-18 & 1.90 & 60,500 & 350.7 & 1.04$^\circ$\\
\hline
\end{tabular}
\label{tab:data}
\vspace{.3in}
\end{table}

\subsection{Northern data set}
\begin{figure}[h]
\centering
\begin{minipage}[b]{0.48\textwidth}
    \includegraphics[width=\textwidth]{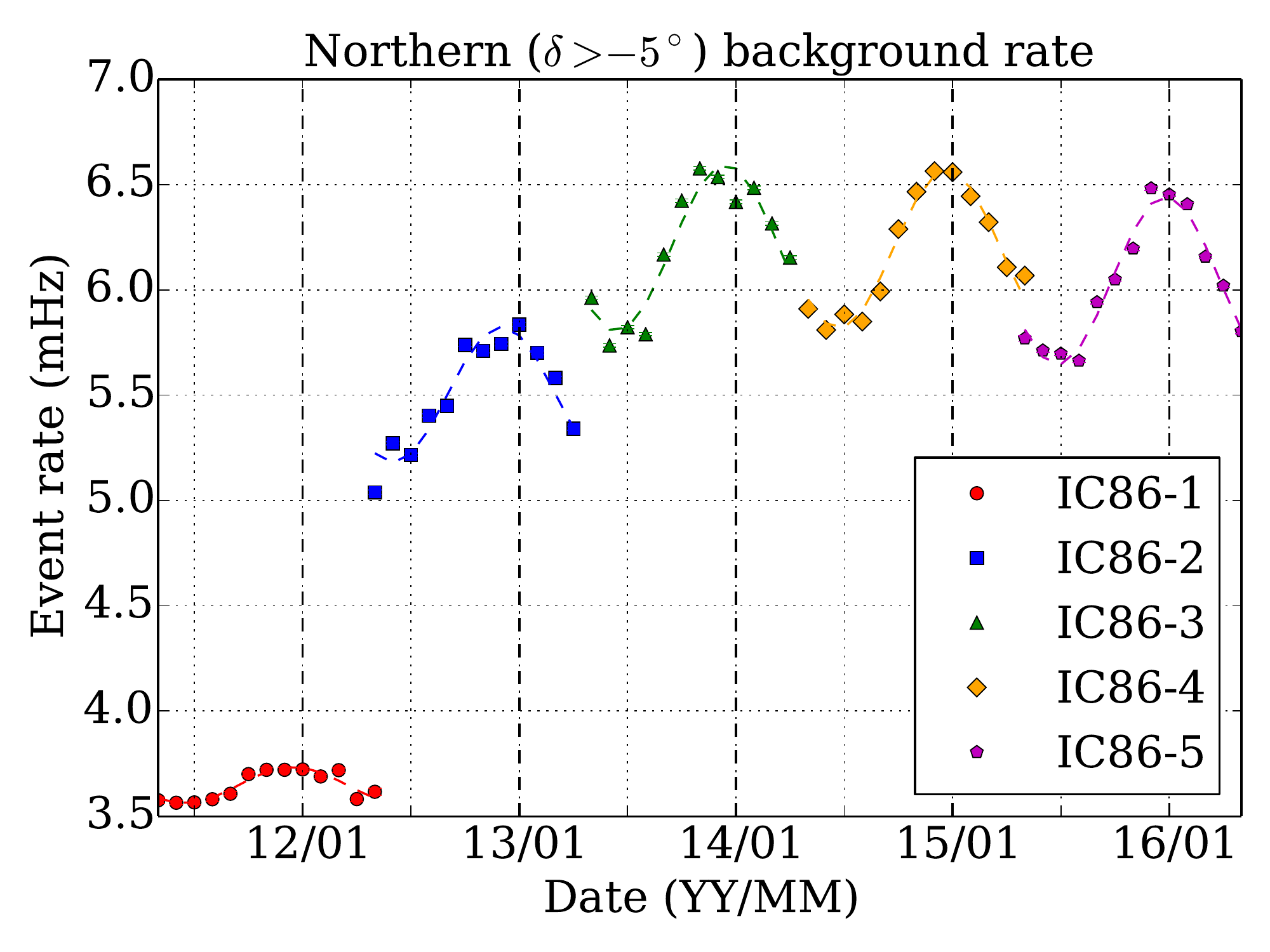}
  \end{minipage}
  \hfill
  \begin{minipage}[b]{0.48\textwidth}
    \includegraphics[width=\textwidth]{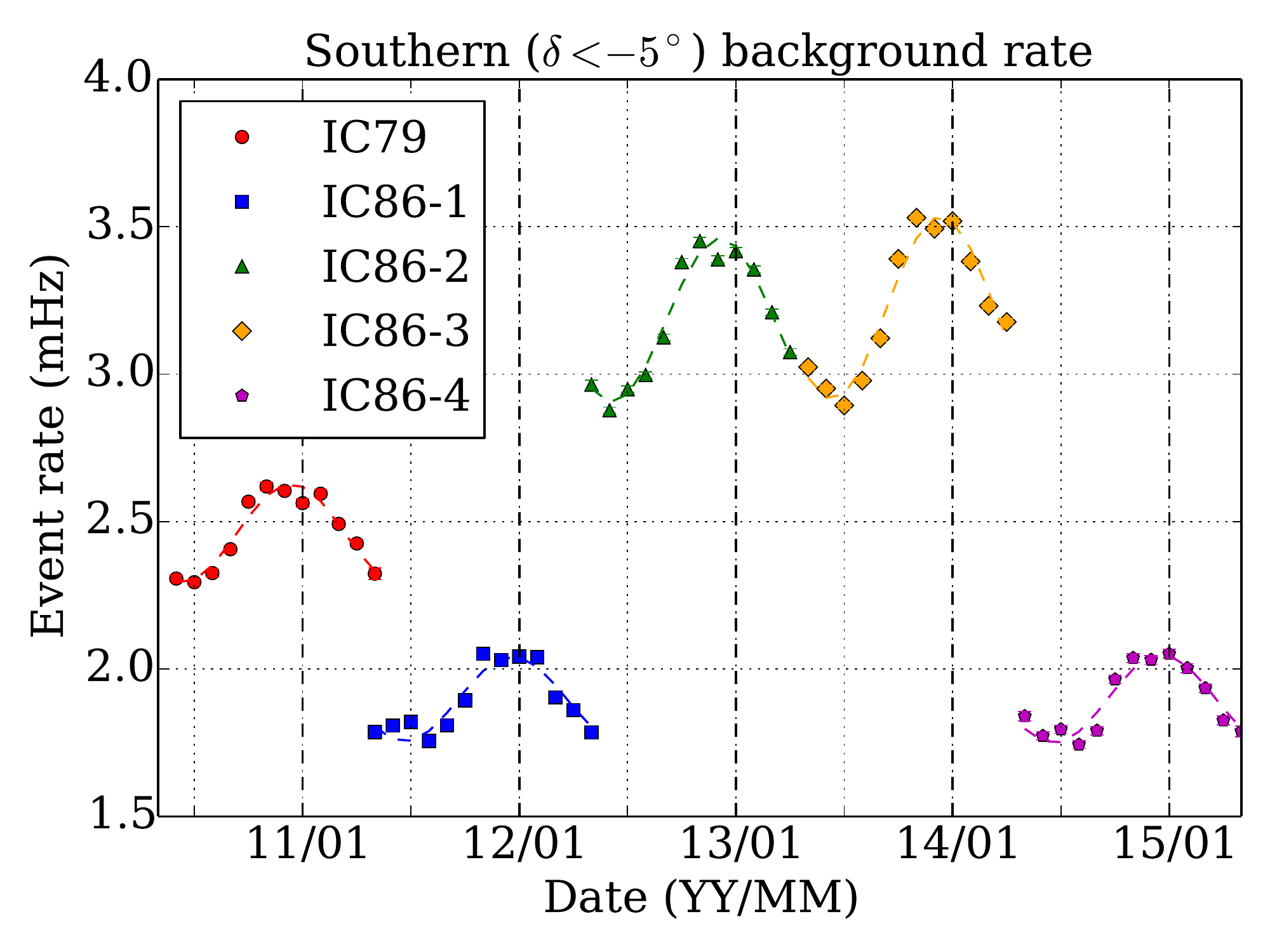}
  \end{minipage}
  \caption{Event rates are shown for each data sample, binned by month and fit yearly with one period of a sine function. Year-to-year rate fluctuations reflect changes in event selection methods, not physical changes to the detector, while seasonal variation within each year is the result of the temperature dependence of atmospheric properties which affect atmospheric muon rates. In the northern hemisphere, seasonal variation accounts for a 2-5\% amplitude (mean-to-peak) variation in the year-averaged rate. In the southern hemisphere, the amplitude of this fluctuation is 7-10\%.}
  \label{fig:seasonalVar}
\end{figure}

The northern data samples ($\delta>-5^\circ$) cover five years of IceCube operation from 2011 May 13 to 2016 May 12, during which 20 northern FRBs were detected (Table~\ref{tab:FRBs}): three each from a unique source and 17 bursts from FRB 121102. In the northern hemisphere, the Earth filters out cosmic ray-induced atmospheric muons, so the data samples consist primarily of atmospheric muon neutrinos with a median energy on the order of 1 TeV. The event rate in the northern hemisphere increases from 3.5 mHz in the first year \citep{aartsen2015search} to 6 mHz in later years \citep{Aartsen:2017wea}, as shown in Figure~\ref{fig:seasonalVar}. This year-to-year variation is due largely to two combined effects: first, the initial event selections treat each year of the IceCube data sample independently due to filter and data processing scheme updates in the early years of IceCube operation; second, each data sample was separately optimized for sensitivity to its corresponding set of GRBs\footnote{In the northern data set, the IC86-1 sample was optimized for sensitivity to a \emph{stacking} search for GRBs. In later years, sensitivity to a \emph{max-burst} search was instead optimized, accounting for the large year-to-year rate fluctuation between samples IC86-1 and IC86-2 (see Figure~\ref{fig:seasonalVar}, Table~\ref{tab:data}).}. 

\begin{figure}[h]
\centering
\begin{minipage}[b]{0.48\textwidth}
    \includegraphics[width=\textwidth]{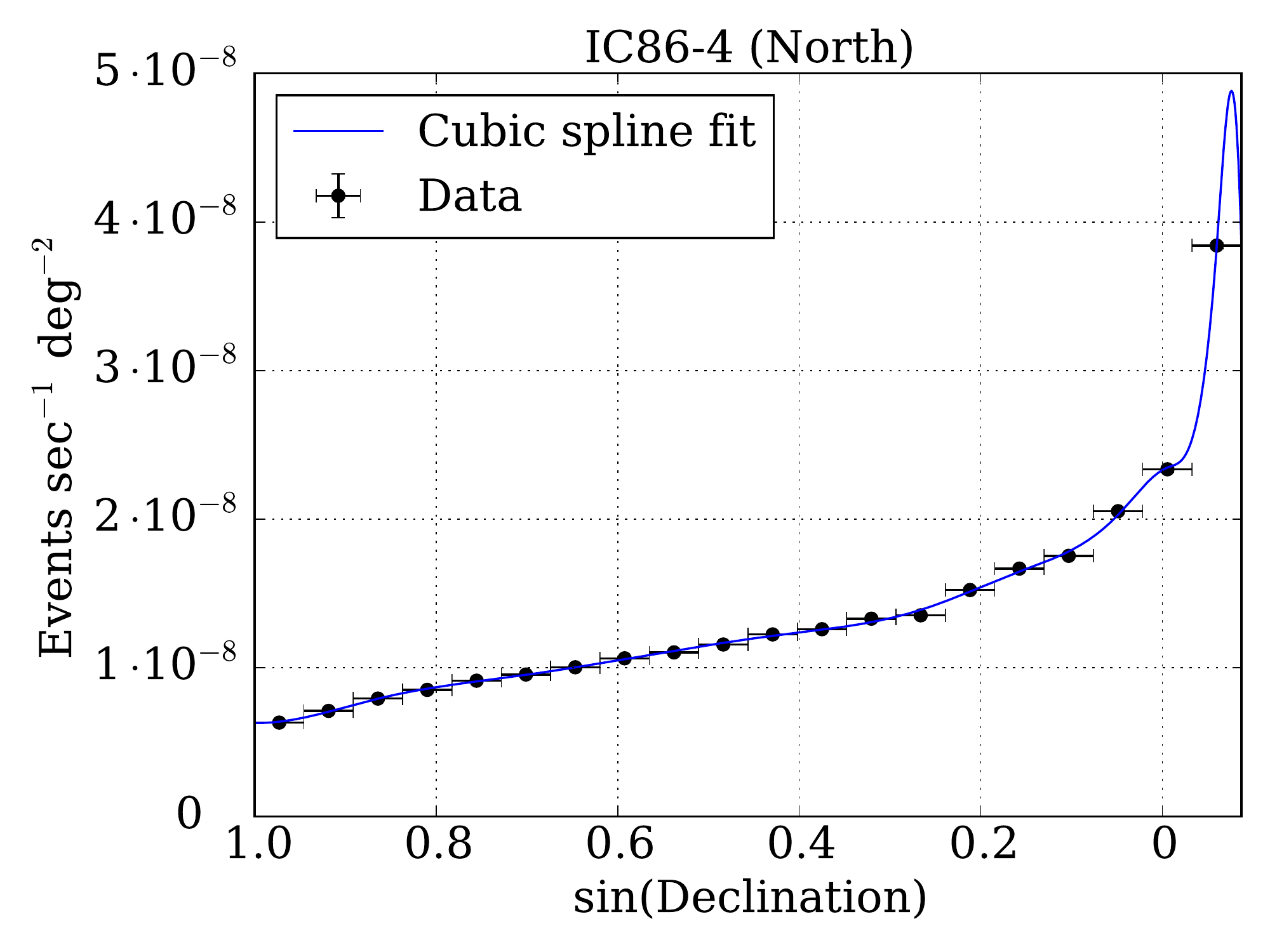}
  \end{minipage}
  \hfill
  \begin{minipage}[b]{0.48\textwidth}
    \includegraphics[width=\textwidth]{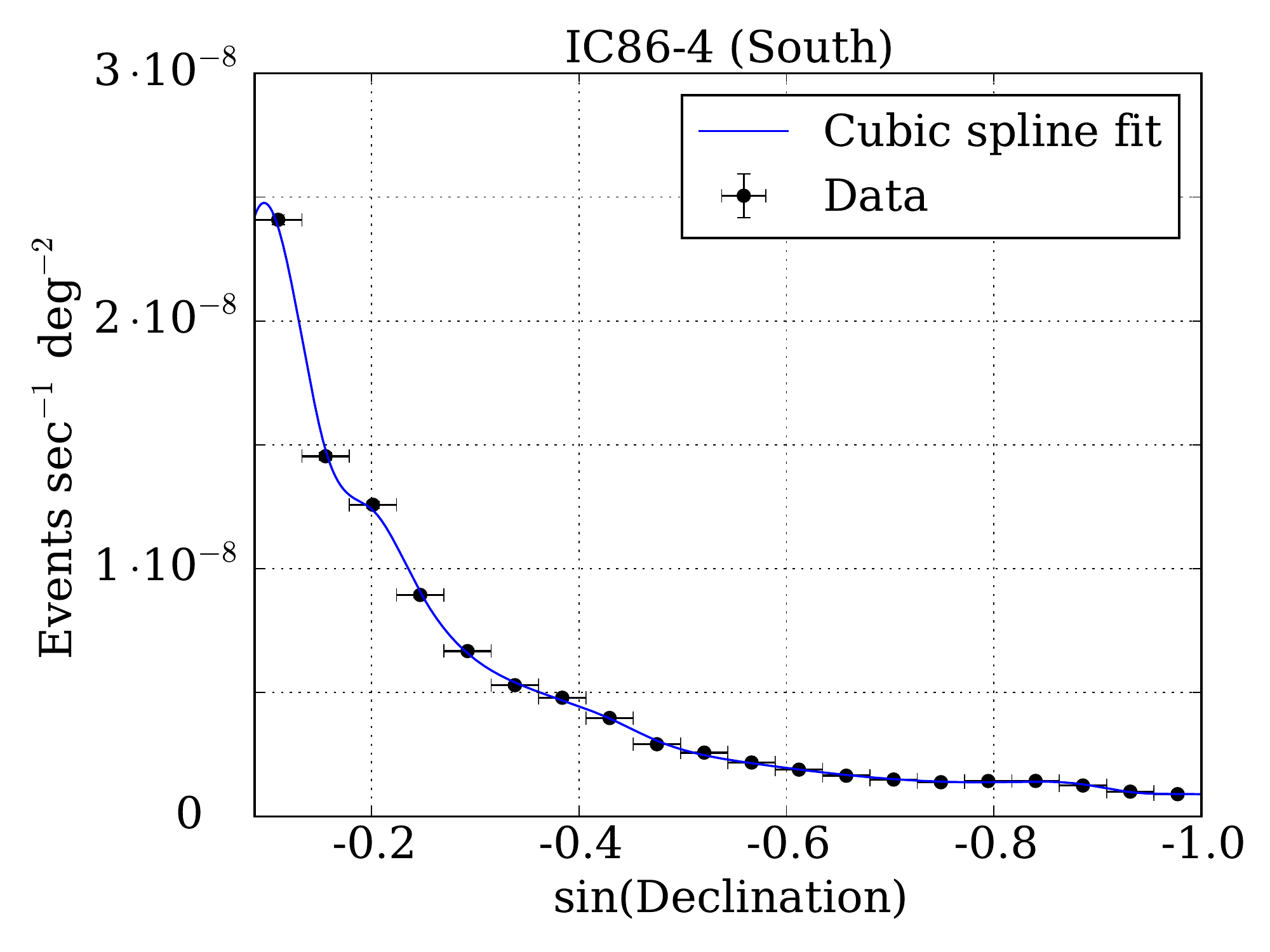}
  \end{minipage}
  \caption{The distribution of events in reconstructed declination is shown for a representative year of off-time data in each hemisphere. Data samples are binned into 20 bins of equal width in sin(declination) and fit with a cubic polynomial spline with endpoints equal to the first and last bin values.}
  \label{fig:zenPDF}
\end{figure}

Within each year, a seasonal variation of the background rate can also be seen \citep{Aartsen:2013jla}. In the Austral summer, the warming atmosphere expands and increases the average height and mean free path of products from cosmic-ray interactions, allowing pions to more frequently decay into $\mu+\nu_{\mu}\footnote{IceCube cannot differentiate between neutrinos and anti-neutrinos, so here $\nu_{\mu}$ denotes both neutrinos and anti-neutrinos}$ and increasing the overall rate of atmospheric muons and neutrinos in IceCube. The phase of the seasonal variation in the northern sample is the same as that in the southern sample because the northern sample is dominated by events between $+15^\circ$ and $-5^\circ$ in declination (Figure~\ref{fig:zenPDF}), which corresponds to production in the atmosphere at latitudes between $-60^\circ$ and $-90^\circ$.

\subsection{Southern data set}
\begin{figure}[h]
\centering
\begin{minipage}[b]{0.48\textwidth}
    \includegraphics[width=\textwidth]{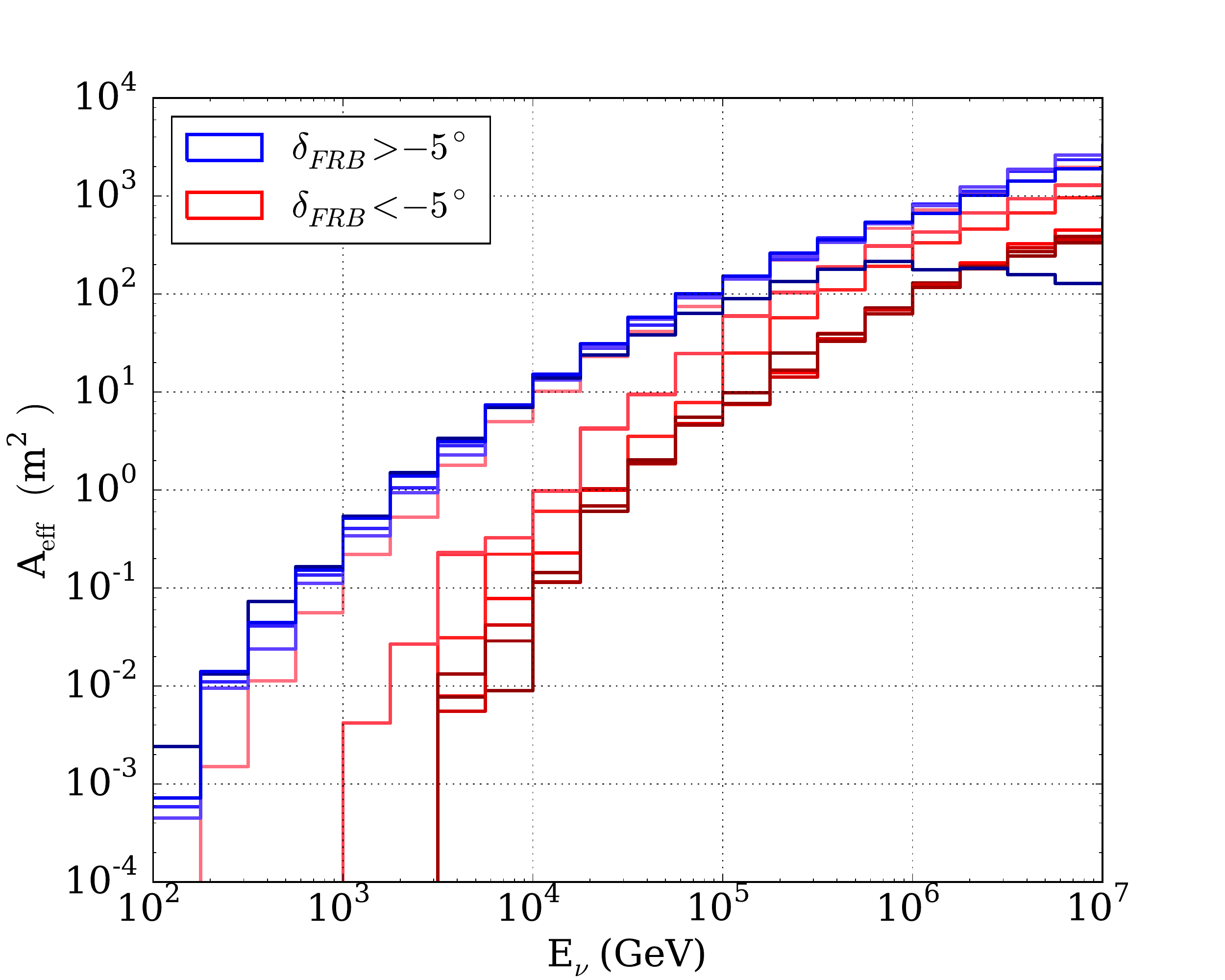}
\end{minipage}
\hfill
\begin{minipage}[b]{0.5\textwidth}
    \includegraphics[width=\textwidth]{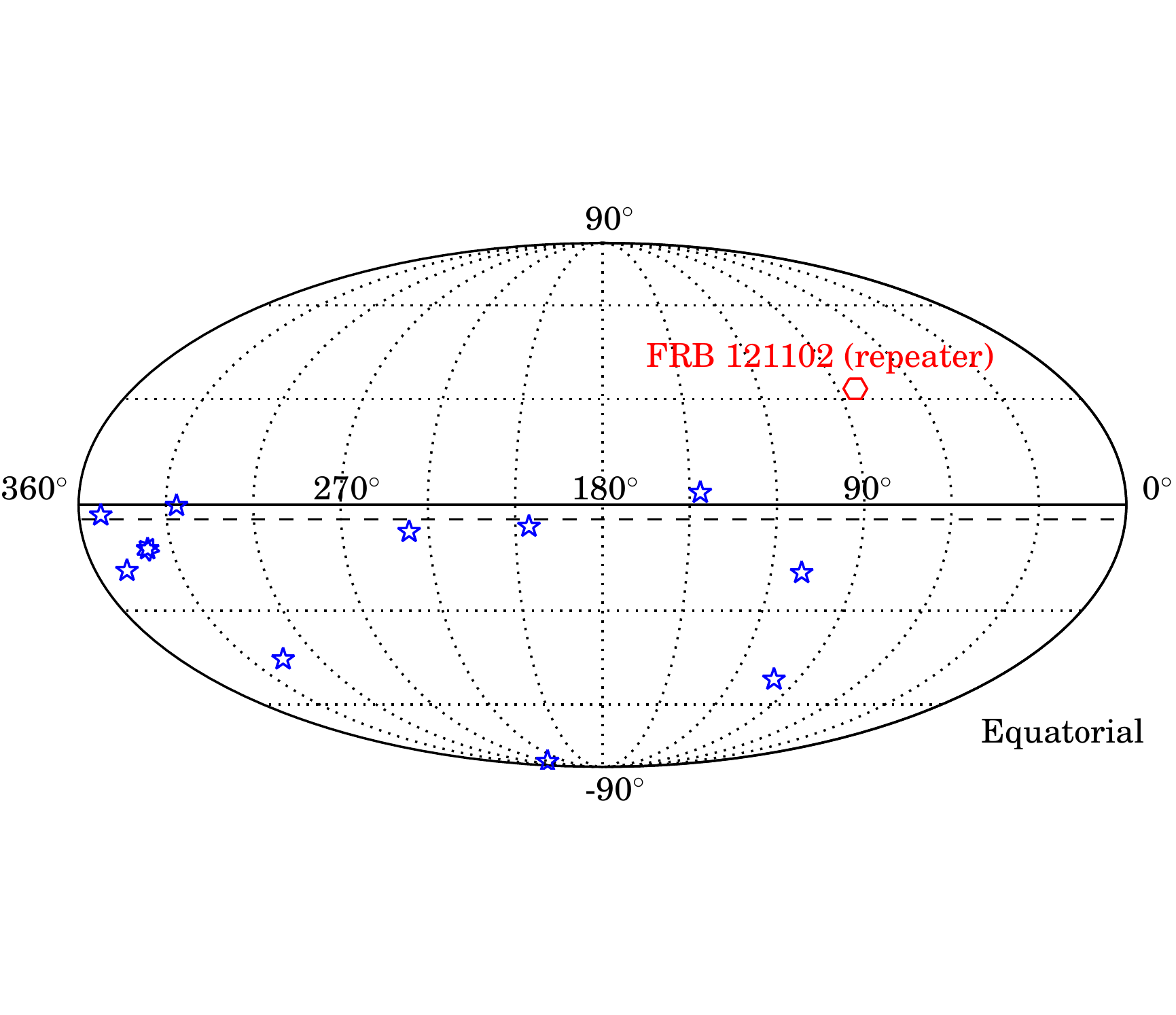}
\end{minipage}
\caption{{\it Left}: the effective area of IceCube to muon neutrinos with energies 100 GeV - 10 PeV is shown for the event selection applied to this analysis' data samples. The effective area was calculated for a declination range ${\Delta \sin(\delta) = 0.04}$ centered on the declination of each FRB using the event selection corresponding each FRB's respective year of data. In total, 15 unique curves are plotted (three curves are calculated for FRB 121102, one for each year during which it was detected), although they sometimes overlap. Northern- and southern-sky FRBs are plotted in blue and red respectively, with a color scale for each corresponding to declination (darker near poles and lighter near celestial equator). In the southern hemisphere, IceCube's energy cuts to reduce atmospheric muon contamination result in a smaller effective area at lower energies. In the northern hemisphere, the effective area of IceCube benefits from shielding by the Earth from muons until, at energies above 100 TeV, the increased neutrino-nucleon cross section results in significant absorption of up-going neutrinos. However, because of the declinations of these FRBs, this effect is only easily seen here for FRB 121102 (${\delta=33^{\circ}}$, dark blue curve), for which $A_{\mathrm{eff}}$ begins to decrease at 1 PeV. {\it Right}: FRB locations in the sky. The FRB 121102 (red hexagon) has repeated 16 times, and the other FRBs (blue stars) have not been observed to repeat.}
\label{fig:Aeff_FRB}
\end{figure}

The southern data samples ($\delta<-5^\circ$) consist of five years of data from 2010 May 31 to 2015 May 18, during which nine southern FRBs were detected. The year-to-year event rate, 2-3.5 mHz, is lower than that of the northern samples due mainly to a higher energy threshold imposed to reduce background from atmospheric muons and the asymmetric separation of hemispheres which makes the northern hemisphere \textasciitilde20\% larger in solid angle than the southern~\citep{Aartsen:2017wea}. The southern samples are dominated by down-going atmospheric muons with median energy on the order of 10 TeV. The effective area of IceCube to neutrino events which pass the event selection can be seen in Figure~\ref{fig:Aeff_FRB}, where the effective area has been determined for the declination of each FRB in this analysis.

\begin{table}[h]
\caption{29 FRBs are included in this search: in the north, 20 bursts from 4 unique source locations, and in the south, 9 bursts each with a unique location. For each FRB, arrival time and dispersion-measure-corrected burst duration are provided with RA and Dec (J2000), as well as the IceCube data sample being recorded during its detection. For FRB 121102, which has been found to repeat, we label individual bursts with "b0", "b1", etc., sorted chronologically by time of detection. FRB 121002 was detected as two bursts separated by \textasciitilde1 ms. It is treated as a single burst in this analysis, but we give both burst durations for completeness.}
\vspace{.1in}
\centering
\begin{tabular}{ c c c c c c }
\hline
\hline
Northern ($\delta>-5^\circ$) FRBs & Time (UTC) & Duration (ms) & RA & Dec & IceCube Data Sample \\
\hline
\hline
FRB 110523 & 2011-05-23 15:06:19.738 & 1.73 & 21h 45$^\prime$ & -00$^\circ$ 12$^\prime$ & IC86-1 \\
\hline
FRB 110703 & 2011-07-03 18:59:40.591 & $<4.3$ & 23h 30$^\prime$ & -02$^\circ$ 52$^\prime$ & IC86-1 \\
\hline
FRB 121102 b0& 2012-11-02 06:47:17.117 & 3.3 & 05h 32$^\prime$ & 33$^\circ$ 05$^\prime$ & IC86-2 \\
\hline
FRB 130628 & 2013-06-28 03:58:00.02 & $<0.05$ & 09h 03$^\prime$ & 03$^\circ$ 26$^\prime$ & IC86-3 \\
\hline
FRB 121102 b1& 2015-05-17 17:42:08.712 & 3.8 & 05h 32$^\prime$ & 33$^\circ$ 05$^\prime$ & IC86-4 \\
\hline
FRB 121102 b2& 2015-05-17 17:51:40.921 & 3.3 & 05h 32$^\prime$ & 33$^\circ$ 05$^\prime$ & IC86-4 \\
\hline
FRB 121102 b3& 2015-06-02 16:38:07.575 & 4.6 & 05h 32$^\prime$ & 33$^\circ$ 05$^\prime$ & IC86-5 \\
\hline
FRB 121102 b4& 2015-06-02 16:47:36.484 & 8.7 & 05h 32$^\prime$ & 33$^\circ$ 05$^\prime$ & IC86-5 \\
\hline
FRB 121102 b5& 2015-06-02 17:49:18.627 & 2.8 & 05h 32$^\prime$ & 33$^\circ$ 05$^\prime$ & IC86-5 \\
\hline
FRB 121102 b6& 2015-06-02 17:49:41.319 & 6.1 & 05h 32$^\prime$ & 33$^\circ$ 05$^\prime$ & IC86-5 \\
\hline
FRB 121102 b7& 2015-06-02 17:50:39.298 & 6.6 & 05h 32$^\prime$ & 33$^\circ$ 05$^\prime$ & IC86-5 \\
\hline
FRB 121102 b8& 2015-06-02 17:53:45.528 & 6.0 & 05h 32$^\prime$ & 33$^\circ$ 05$^\prime$ & IC86-5 \\
\hline
FRB 121102 b9& 2015-06-02 17:56:34.787 & 8.0 & 05h 32$^\prime$ & 33$^\circ$ 05$^\prime$ & IC86-5 \\
\hline
FRB 121102 b10& 2015-06-02 17:57:32.020 & 3.1 & 05h 32$^\prime$ & 33$^\circ$ 05$^\prime$ & IC86-5 \\
\hline
FRB 121102 b11& 2015-11-13 08:32:42.375 & 6.73 & 05h 32$^\prime$ & 33$^\circ$ 05$^\prime$ & IC86-5 \\
\hline
FRB 121102 b12& 2015-11-19 10:44:40.524 & 6.10 & 05h 32$^\prime$ & 33$^\circ$ 05$^\prime$ & IC86-5 \\
\hline
FRB 121102 b13& 2015-11-19 10:51:34.957 & 6.14 & 05h 32$^\prime$ & 33$^\circ$ 05$^\prime$ & IC86-5 \\
\hline
FRB 121102 b14& 2015-11-19 10:58:56.234 & 4.30 & 05h 32$^\prime$ & 33$^\circ$ 05$^\prime$ & IC86-5 \\
\hline
FRB 121102 b15& 2015-11-19 11:05:52.492 & 5.97 & 05h 32$^\prime$ & 33$^\circ$ 05$^\prime$ & IC86-5 \\
\hline
FRB 121102 b16& 2015-12-08 04:54:40.262 & 2.50 & 05h 32$^\prime$ & 33$^\circ$ 05$^\prime$ & IC86-5 \\
\hline
\hline
Southern ($\delta<-5^\circ$) FRBs & Time (UTC) & Duration (ms) & RA & Dec & IceCube Data Sample \\
\hline
\hline
FRB 110220 & 2011-02-20 01:55:48.957 & 5.6 & 22h 34$^\prime$ & -12$^\circ$ 24$^\prime$ & IC79 \\
\hline
FRB 110627 & 2011-06-27 21:33:17.474 & $<1.4$ & 21h 03$^\prime$ & -44$^\circ$ 44$^\prime$ & IC86-1 \\
\hline
FRB 120127 & 2012-01-27 08:11:21.723 & $<1.1$ & 23h 15$^\prime$ & -18$^\circ$ 25$^\prime$ & IC86-1 \\
\hline
FRB 121002 & 2012-10-02 13:09:18.402 & 2.1; 3.7 & 18h 14$^\prime$ & -85$^\circ$ 11$^\prime$ & IC86-2 \\
\hline
FRB 130626 & 2013-06-26 14:56:00.06 & $<0.12$ & 16h 27$^\prime$ & -07$^\circ$ 27$^\prime$ & IC86-3 \\
\hline
FRB 130729 & 2013-07-29 09:01:52.64 & $<4$ & 13h 41$^\prime$ & -05$^\circ$ 59$^\prime$ & IC86-3 \\
\hline
FRB 131104 & 2013-11-04 18:04:01.2 & $<0.64$ & 06h 44$^\prime$ & -51$^\circ$ 17$^\prime$ & IC86-3 \\
\hline
FRB 140514 & 2014-05-14 17:14:11.06 & 2.8 & 22h 34$^\prime$ & -12$^\circ$ 18$^\prime$ & IC86-4 \\
\hline
FRB 150418 & 2015-04-18 04:29:05.370 & 0.8 & 07h 16$^\prime$ & -19$^\circ$ 00$^\prime$ & IC86-4 \\
\hline
\end{tabular}
\label{tab:FRBs}
\vspace{.1in}
\end{table}      

\newpage
\section{Analysis Methods} \label{ana_method}

\subsection{Unbinned likelihood method}

An unbinned maximum likelihood method is used to search for spatial and temporal coincidence of neutrino events with detected FRBs \citep{Aartsen:2014aqy}. In a given coincidence window $\Delta$T centered on the time of detection of each FRB, the likelihood of observing $N$ events for an expected $(n_s+n_b)$ events is
\begin{equation}
\mathscr{L}(N,\{x_i\};n_s+n_b)=\frac{(n_s+n_b)^N}{N!}\cdot \exp{ \lbrack -(n_s+n_b) \rbrack}\cdot \prod_{i=1}^{N}\frac{n_s S(x_i)+n_b B(x_i)}{n_s+n_b}
\end{equation}
where $n_s$ and $n_b$ are the expected number of observed signal and background events, $x_i$ is the reconstructed direction and estimated angular uncertainty for each event $i$, $S(x_i)$ is the signal PDF -- taken to be a radially symmetric 2D Gaussian with standard deviation $\sigma_{i}$ -- evaluated for the angular separation between event $i$ and the FRB with which it is temporally coincident, and $B(x_i)$ is the background PDF for the data sample to which event $i$ belongs evaluated at the declination of event $i$. The uncertainties of the FRB locations are taken into account in $S(x_i)$, but they are significantly smaller than the median angular uncertainty of the data. In any time window $\Delta$T, the $N$ events are those which IceCube detected within $\pm\Delta$T$/2$ of any FRB detection. Before background event rates and PDFs were calculated, \emph{on-time data} -- data collected within $\pm 2$ days of any FRB detection -- were removed from the samples until all analysis procedures were determined. The remaining data (>1700 days of data per hemisphere) are considered \emph{off-time data}, which we used to determine background characteristics to prevent artificial bias from affecting the results of our search. Figure~\ref{fig:zenPDF} shows examples of off-time data distributions for both northern and southern hemispheres.

A generic test statistic (TS) is used in this analysis, defined as the logarithmic ratio of the likelihood of the alternative hypothesis $\mathscr{L} (N,\{x_i\}; n_s+n_b)$ and that of the null hypothesis $\mathscr{L}_0 (N,\{x_i\}; n_b)$, which can be written as
\begin{equation}
\textrm{TS}= -{n_s}+\sum_{i=1}^{N}\ln\Big[1+\frac{{n_s}S(x_i)}{n_b B(x_i)}\Big]
\label{TS_def}
\end{equation}
The TS is maximized with respect to ${n_s}$ to find the most probable number of signal-like events among $N$ temporally coincident events. $n_b$ is calculated by multiplying time-dependent background rate for each FRB, modeled from off-time data, by $\Delta$T.

Two search strategies are implemented based on this test statistic. The \emph{stacking} search tests the hypothesis that the astrophysical class of FRBs emits neutrinos. In this search, $n_s$ and $n_b$ are the total number of expected signal and background events contained in the time windows of an entire list of FRBs for the hemisphere. One TS value (with its corresponding $n_s$) is returned for an ensemble of $N$ events which consist of on-time events from all the bursts. This TS represents the significance of correlation between the events analyzed and the source class as a whole.
The \emph{max-burst} search tests the hypothesis that one or a few bright sources emit neutrinos regardless of source classification. In this search, $n_s$ and $n_b$ are evaluated separately for each FRB. A TS-$n_s$ pair is calculated for each FRB considering only the events coincident with its time window. The most statistically significant of these TS (and its corresponding $n_s$) is returned as the max-burst TS value of the ensemble.

Since neutrino emission mechanisms and potential neutrino arrival times relative to the time of radio detection are unknown, we employ a model-independent search using an expanding time window, similar to a previous search for prompt neutrino emission from gamma ray bursts by IceCube which found no correlation \citep{Abbasi:2012zw}. Starting with ${\Delta\textrm{T}=0.01~\textrm{s}}$ centered on each FRB, we search a series of time windows expanding by factors of two, i.e. ${\Delta\textrm{T}=2^j \cdot (0.01~\textrm{s})}$ for ${j=0,1,2, ... , 24}$. We stop expanding at a time window size of 1.94 days (167772.16 s), where the background becomes significant. For the repeating burst FRB 121102 with burst separations less than the largest time window searched, time windows of consecutive bursts stop expanding when otherwise they would overlap. 

In the northern max-burst search, a bright radio burst with a flux of 7.5 Jy detected by the LOFAR radio array \citep{Stewart:2015phy} was included. This LOFAR burst was detected on 2011 December 24 at 04:33 UTC, near the North Celestial Pole (RA~=~22$^h$53$^m$47.1$^s$, DEC~=~+86$^{\circ}$21$\arcmin$46.4$\arcsec$) and lasted \textasciitilde11 minutes. The burst was not consistent with an FRB, so it was not included in the stacking search, during which some degree of uniformity among the stacked source class was required.  

\subsection{Background ensembles}
For each search method and hemisphere, we simulate $10^9$ background-only Monte Carlo pseudo-experiments for every $\Delta$T. This is done by first finding the seasonal variation-adjusted background rate (from Figure~\ref{fig:seasonalVar}) for each FRB in the hemisphere. The product of these rates and $\Delta$T gives a set of mean values for the Poisson distributions from which background events will be drawn. In a single trial, the number of events in the time window of each FRB is randomly drawn, and each event is assigned spatial coordinates which are uniform in detector azimuth and have declination values drawn from the PDFs shown in Figure~\ref{fig:zenPDF}. An angular uncertainty for each event is also randomly assigned from the angular distribution of the off-time data \citep{maunu_phd}. The TS value for the trial is maximized with respect to $n_s$ and the process is repeated for $10^9$ trials, forming a TS distribution for the background-only hypothesis. 

\begin{figure}[h]
\centering
\begin{minipage}[b]{0.48\textwidth}
    \includegraphics[width=\textwidth]{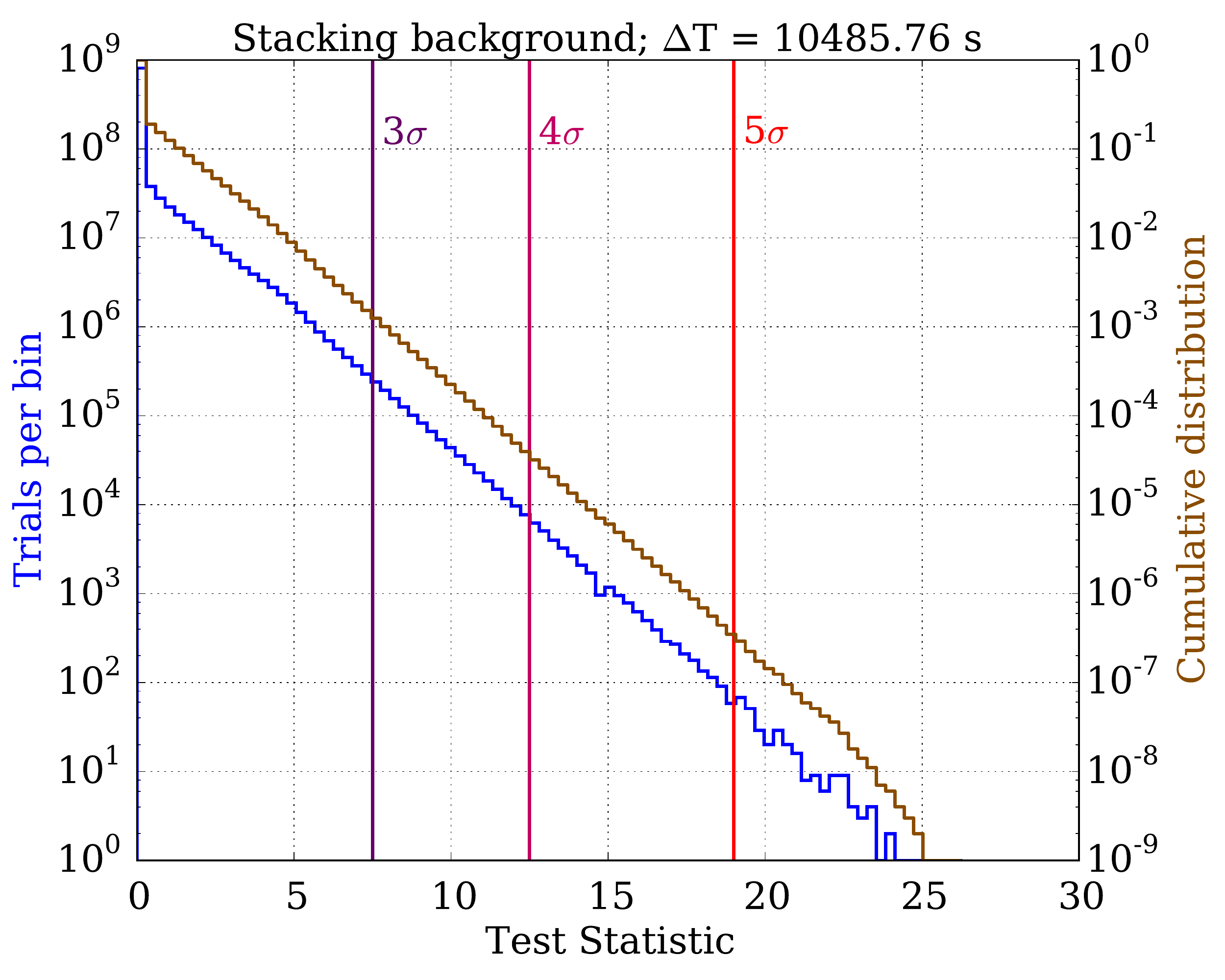}
  \end{minipage}
  \hfill
  \begin{minipage}[b]{0.48\textwidth}
    \includegraphics[width=\textwidth]{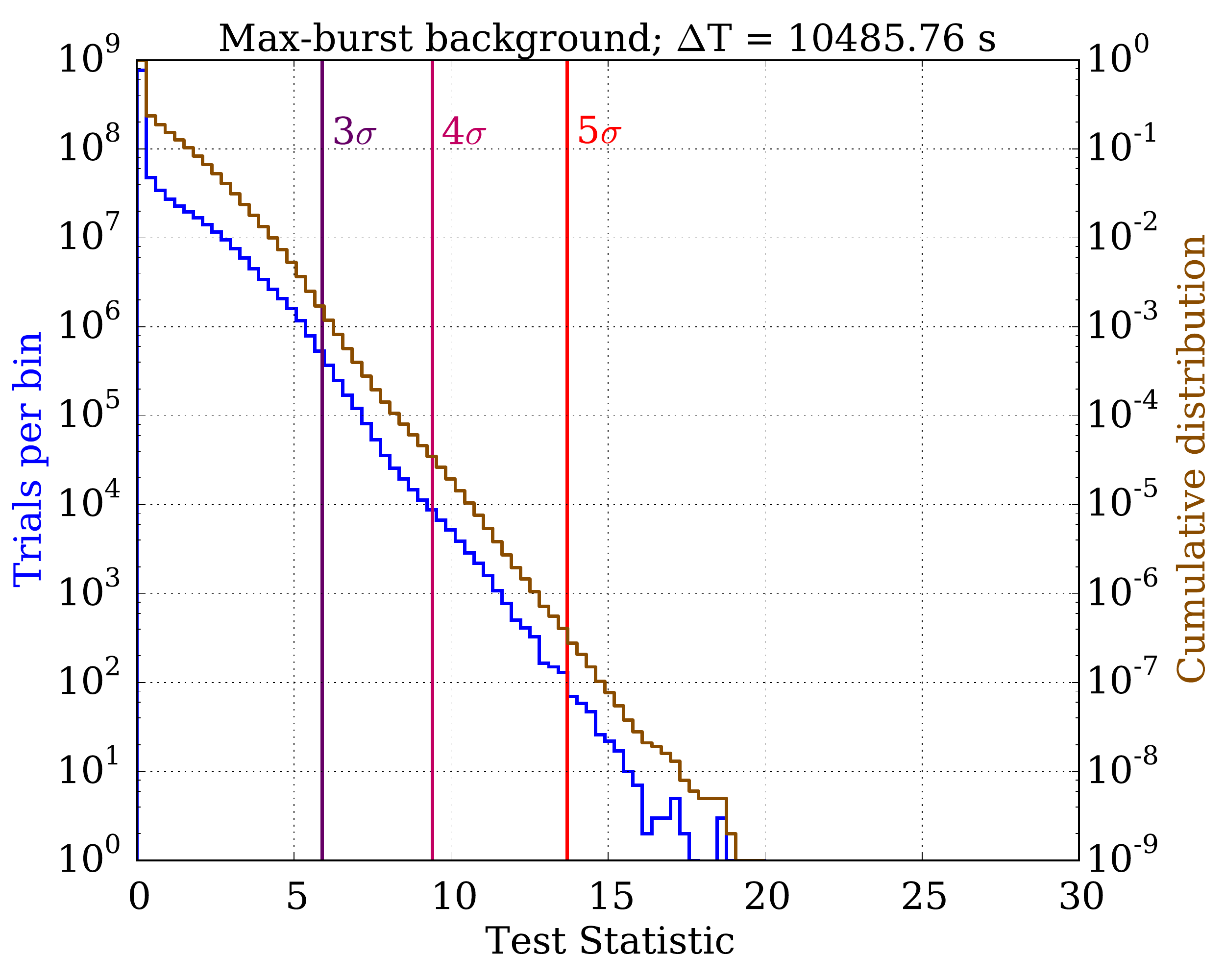}
  \end{minipage}
  \caption{TS distributions are shown for $10^9$ background-only Monte Carlo trials in the southern stacking and max-burst searches at $\Delta$T$~=~10485.76$~s. Significance thresholds (e.g. $5\sigma$) are determined using the corresponding p-value for one tail of a normal distribution. In the low-background regime, each trial is unlikely to contain any spatially coincident events, thus the majority of trials are more background-like than signal-like, returning a negative TS value. These are rounded to zero, resulting in low-background TS distributions peaked sharply at TS$~=~0$. As $\Delta$T increases, the height of the background TS distribution at TS$~=~0$ approaches 50\% of trials as expected.}
  \label{fig:tsd_20}
\end{figure}

For example, Figure~\ref{fig:tsd_20} shows the background-only TS distribution for the southern stacking search at $\Delta\textrm{T}=10485.76~\textrm{s}$. Negative TS values are rounded to zero for the purposes of calculating the significance of analysis results. Building a TS distribution in this manner implicitly factors in a trials factor for the number of bursts searched, since increasing the number of sources inflates the TS values of both the analysis result and the background-only distribution. However, there is an additional trials factor when searching in overlapping time windows, so the cross-time-window trials factor must be accounted for when calculating significance values.  

For each search, the analysis procedure returns the most optimal time window and the corresponding TS-$n_s$ pair, as determined by the p-value of the observed TS in the background-only distribution. Post-trial p-values are obtained by investigating more ensembles of background-only trials. For each trial, a set of events is injected for the largest $\Delta$T following the background-only procedure described above. Then, for each $\Delta$T, a TS value is calculated relative to its corresponding subset of events which are randomly selected from the total event set. The most significant of these TS values has a p-value which becomes one background-only pre-trial p-value. These trials are repeated $10^5$ times, forming a pre-trial p-value distribution. The position of the pre-trial p-value from the search on on-time data in this distribution determines its post-trial p-value.

\section{Sensitivity} \label{sensitivity}

\begin{figure*}[b]
\centering
  \begin{minipage}[b]{0.46\textwidth}
  \centering
  \includegraphics[width=\textwidth]{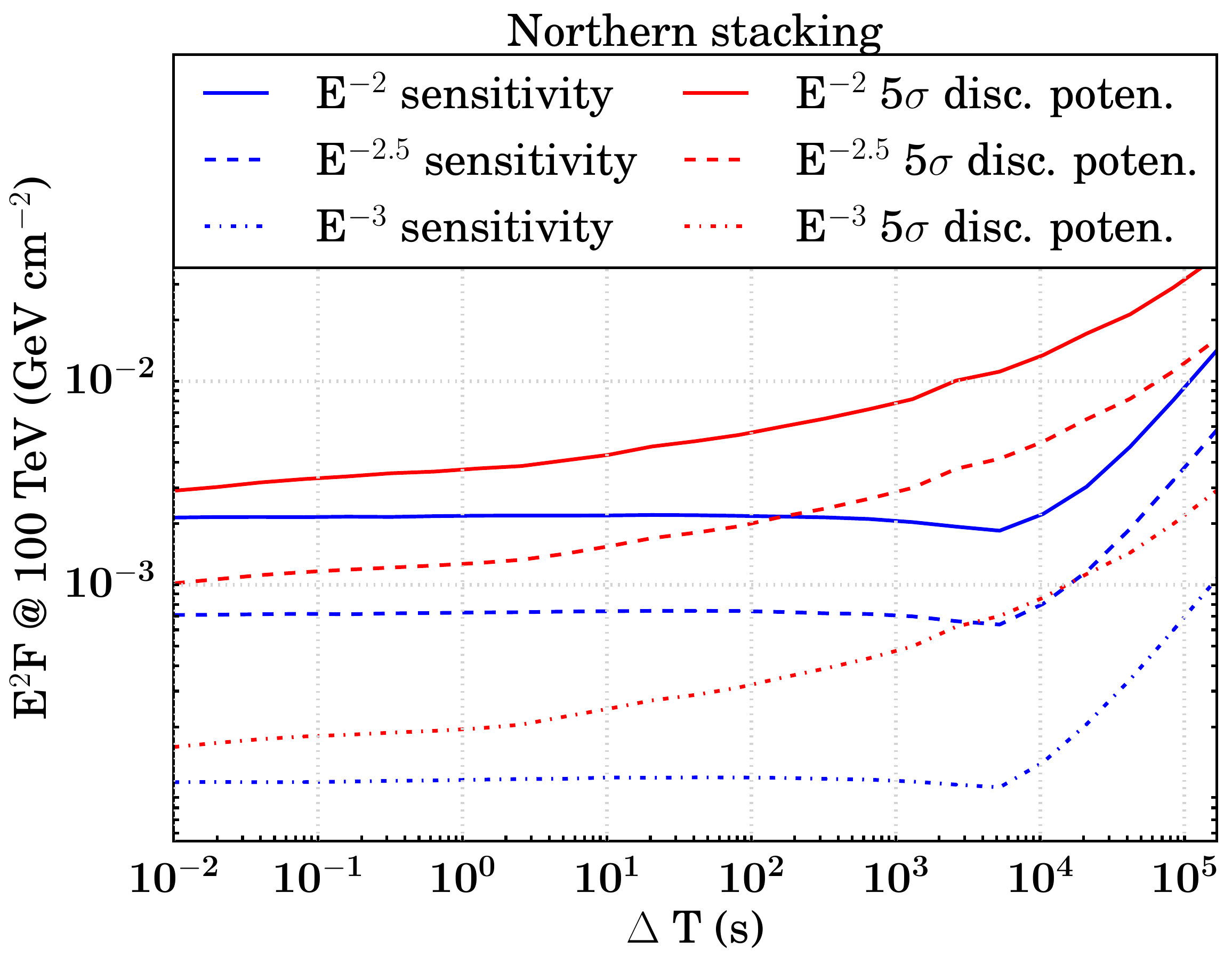}
  \label{fig:sensitivity_N_stacking}
  \end{minipage}
  \quad
  \begin{minipage}[b]{0.46\textwidth}  
  \centering 
  \includegraphics[width=\textwidth]{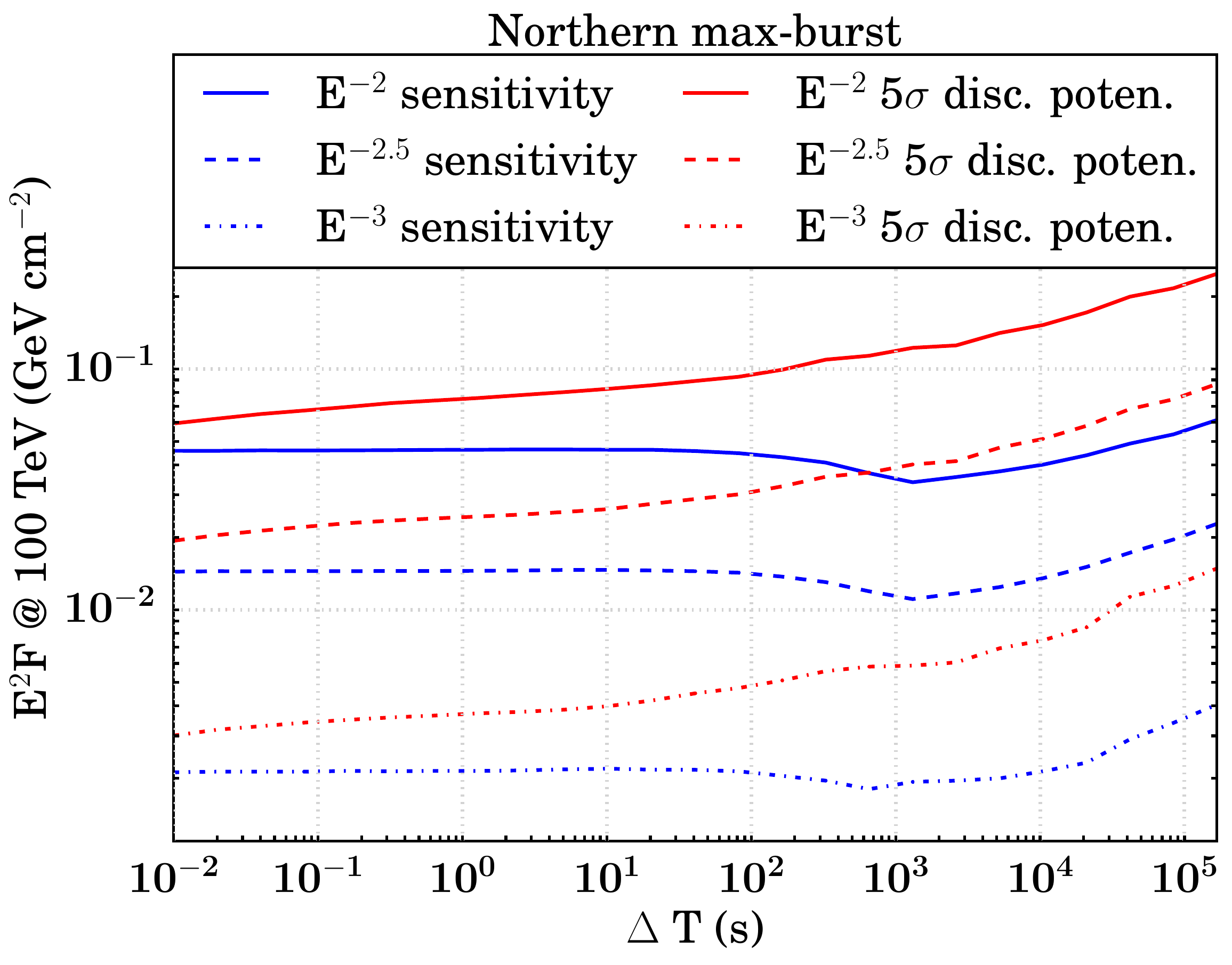} 
  \label{fig:sensitivity_N_maxBurst}
  \end{minipage}
  \quad
  \begin{minipage}[b]{0.46\textwidth}   
  \centering 
  \includegraphics[width=\textwidth]{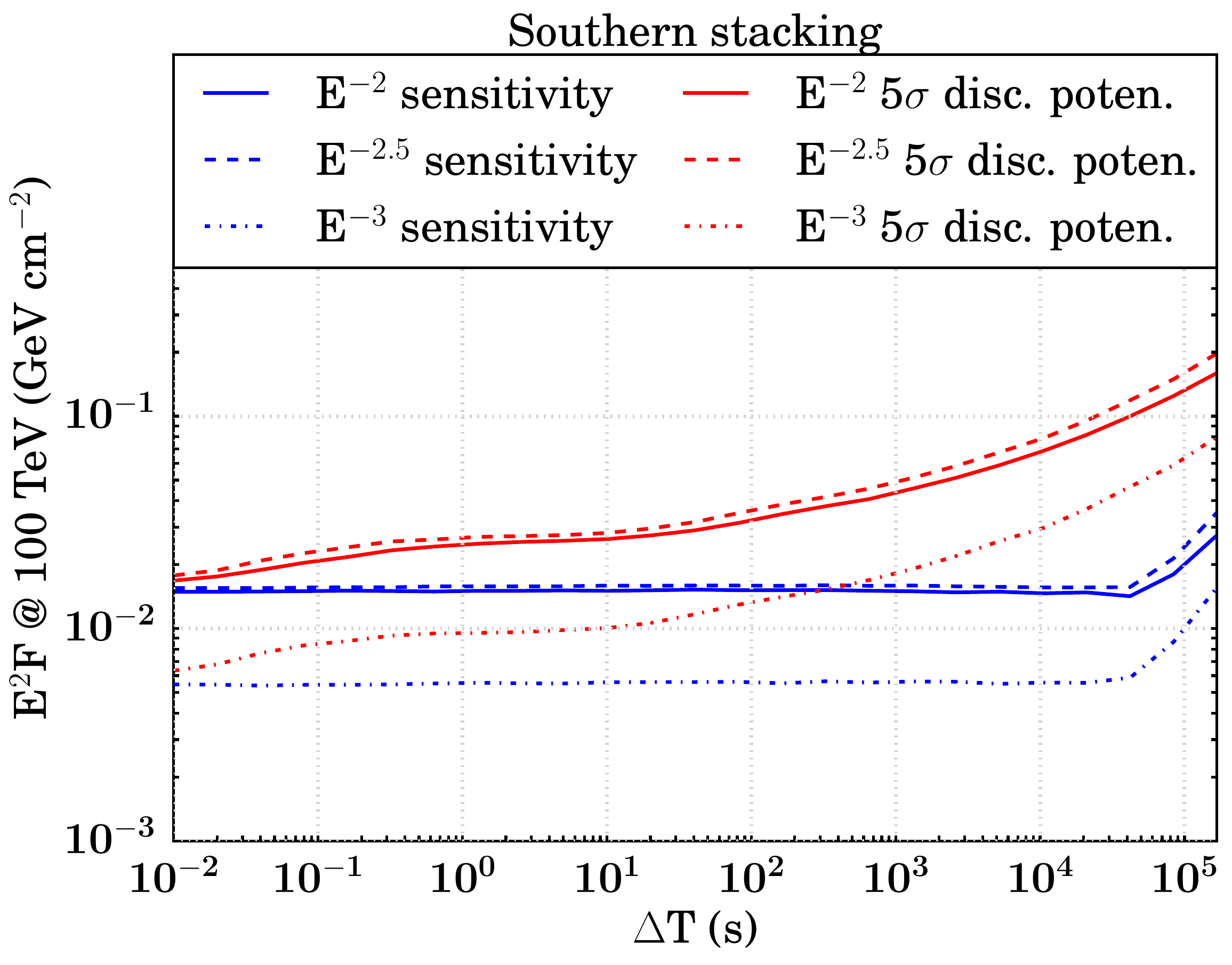}
  \label{fig:sensitivity_S_stacking}
  \end{minipage}
  \quad
  \begin{minipage}[b]{0.46\textwidth}   
  \centering 
  \includegraphics[width=\textwidth]{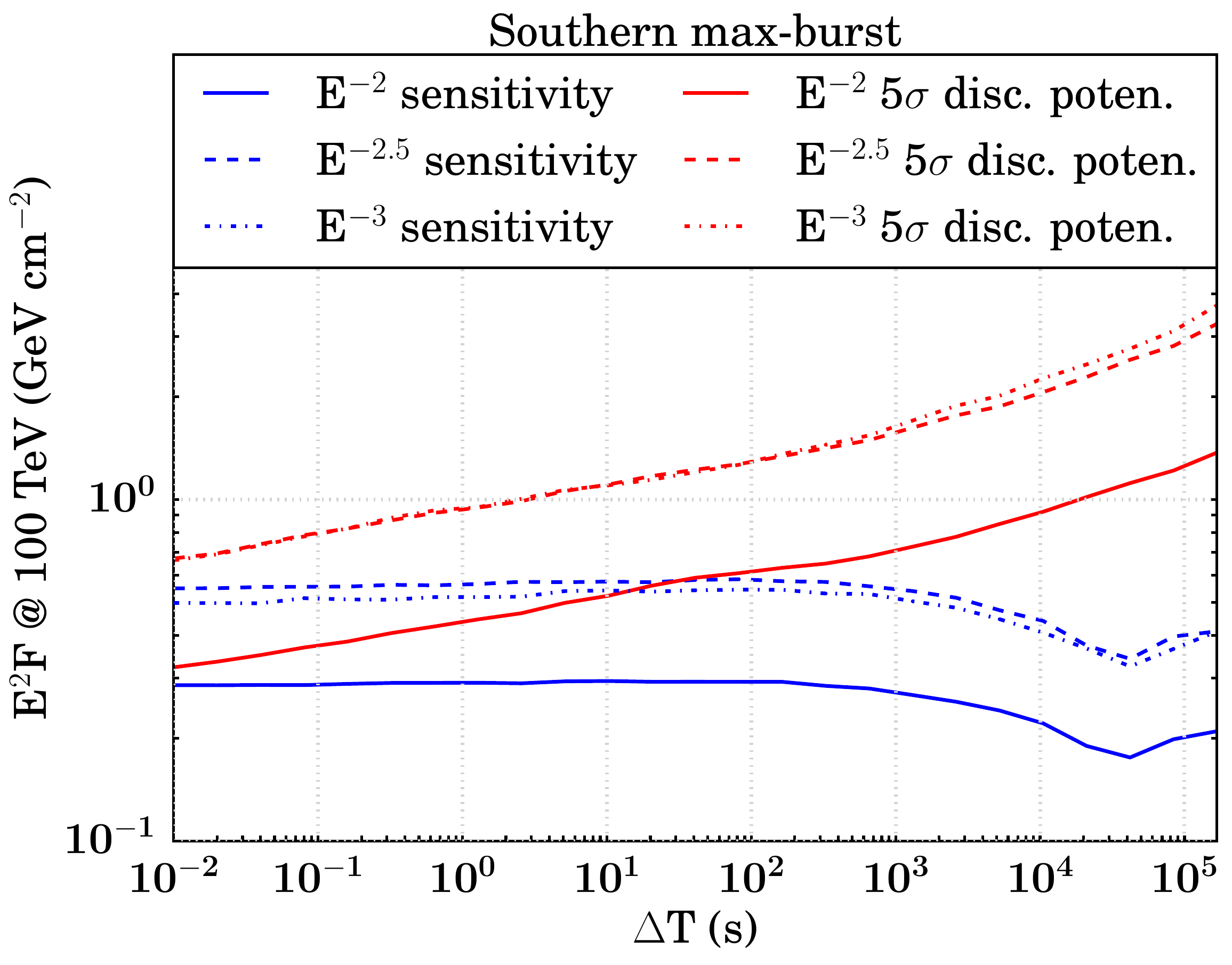}
  \label{fig:sensitivity_S_maxBurst}
  \end{minipage}
  \caption{Sensitivity and 5$\sigma$ discovery potential (90\% confidence level) versus time-window size are shown for the northern hemisphere stacking, northern max-burst, southern stacking, and southern max-burst searches. The values plotted are $E^2$ times the time-integrated flux per burst at 100 TeV, for signal spectra of $E^{-2}$, $E^{-2.5}$, and $E^{-3}$.}
  \label{fig:sensitivity}
\end{figure*}

The sensitivity and discovery potential are calculated by injecting signal events following an assumed unbroken power law energy spectrum ($E^{-2}$, $E^{-2.5}$, and $E^{-3}$) on top of injected background events
. The injected signal fluence (time integrated flux, denoted as $F$) is found which yields a certain probability of obtaining a certain significance in the background-only TS distribution \citep{10.2307/91337, Aartsen:2017wea}. Specifically, sensitivity and discovery potential are defined as the minimum signal fluences required to surpass, respectively, the median in 90$\%$ of the trials and the $5\sigma$ point in 90$\%$ of the trials. 
Figure~\ref{fig:sensitivity} shows the sensitivities and $5\sigma$ discovery potentials for both hemispheres and search strategies. The searches in the northern hemisphere are roughly an order of magnitude more sensitive than those in the south, because of the differences in effective area as described in Section~\ref{event_sample}. 

At $\Delta\textrm{T}=0.01~\textrm{s}$, we expect fewer than 0.001 background events all-hemisphere per trial in each search. As a result, the median background-only TS value is zero for all $\Delta$T until it becomes more probable than not that a background event is injected near an FRB location, resulting in a non-zero TS value. In general, the sensitivity remains constant in a $\Delta$T range that is relatively background-free and transitions to a monotonically increasing function in background-dominated $\Delta$T. We still search all of these low-background $\Delta$T because the discovery potential increases even in the small background regime (Figure~\ref{fig:sensitivity}). 

As a result of our methodology, there is a point in the background transition region where the sensitivity fluence appears to improve. Where the median of the background TS distribution is zero, the 90\% sensitivity threshold for signal injection remains constant. But when $\Delta$T is growing, there are more background events in each trial which can give rise to non-zero TS values, so the injected fluence necessary to meet the criteria for sensitivity is less. Once the median background TS value becomes non-zero, the sensitivity increases as expected.

\section{Results} \label{results}

After correcting for trials factors induced by 25 overlapping time windows searched, no significant correlation between neutrino events and FRBs is found (nor with the LOFAR burst). The most significant pre-trial p-value ($p=0.034$) is found in the northern max-burst search at ${\Delta\textrm{T}=655.36~\textrm{s}}$, with best-fit TS and $n_s$ of 3.90 and 0.99 respectively. The post-trial p-value for this search is $p=0.25$. In the same $\Delta$T, the northern stacking search returned a best-fit TS and $n_s$ of 1.41 and 1.01 respectively, corresponding to a pre-trial p-value $p=0.074$ and post-trial p-value $p=0.375$. The most signal-like event for both searches occurred 200.806~s after FRB 121102 b3, with an angular separation of 2.31$^{\circ}$ and estimated angular uncertainty of 1.31$^{\circ}$.

\begin{figure}
\vspace{.2in}
\centering
\begin{minipage}[b]{0.48\textwidth}
    \includegraphics[width=\textwidth]{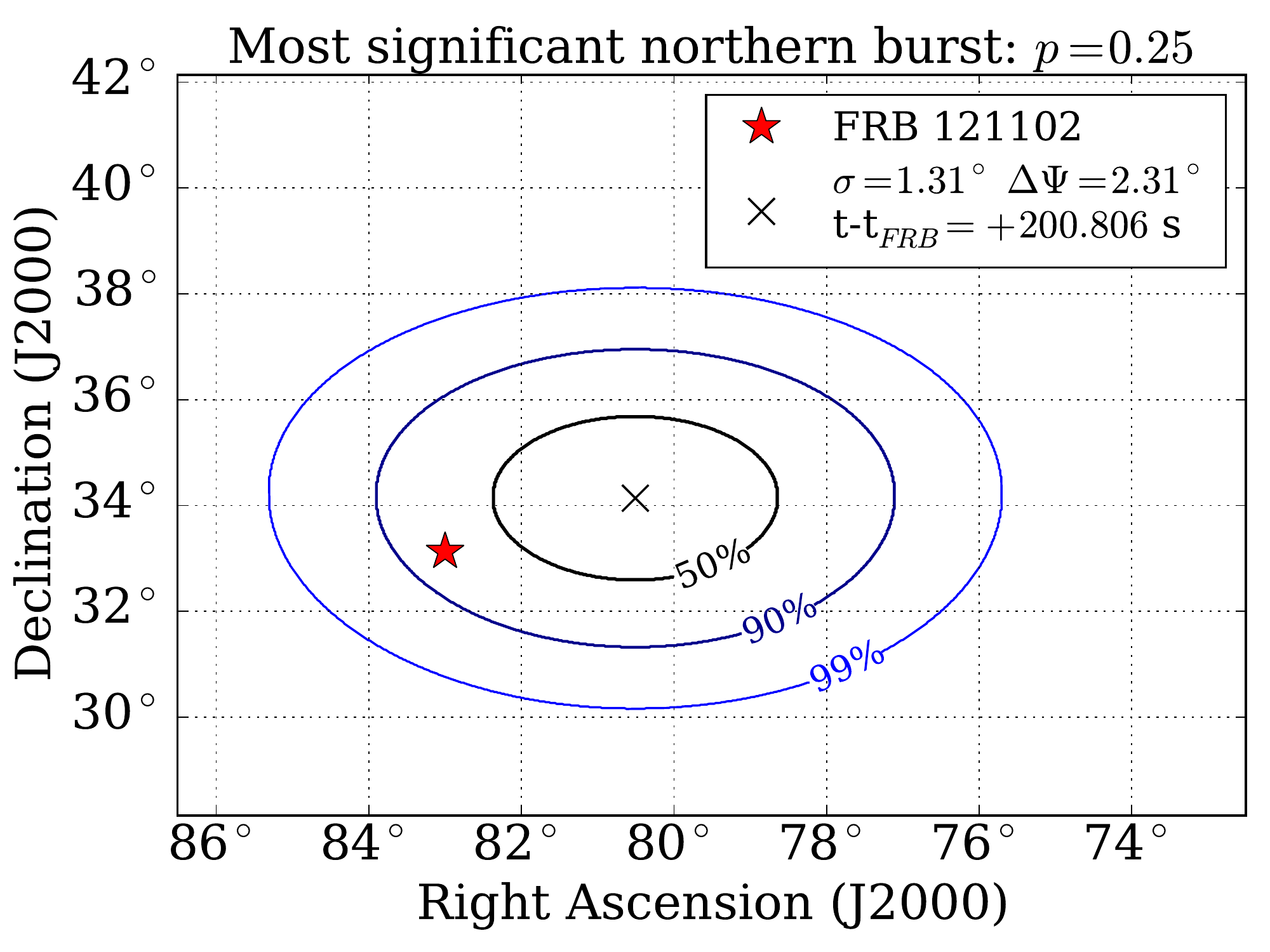}
  \end{minipage}
  \hfill
  \begin{minipage}[b]{0.48\textwidth}
    \includegraphics[width=\textwidth]{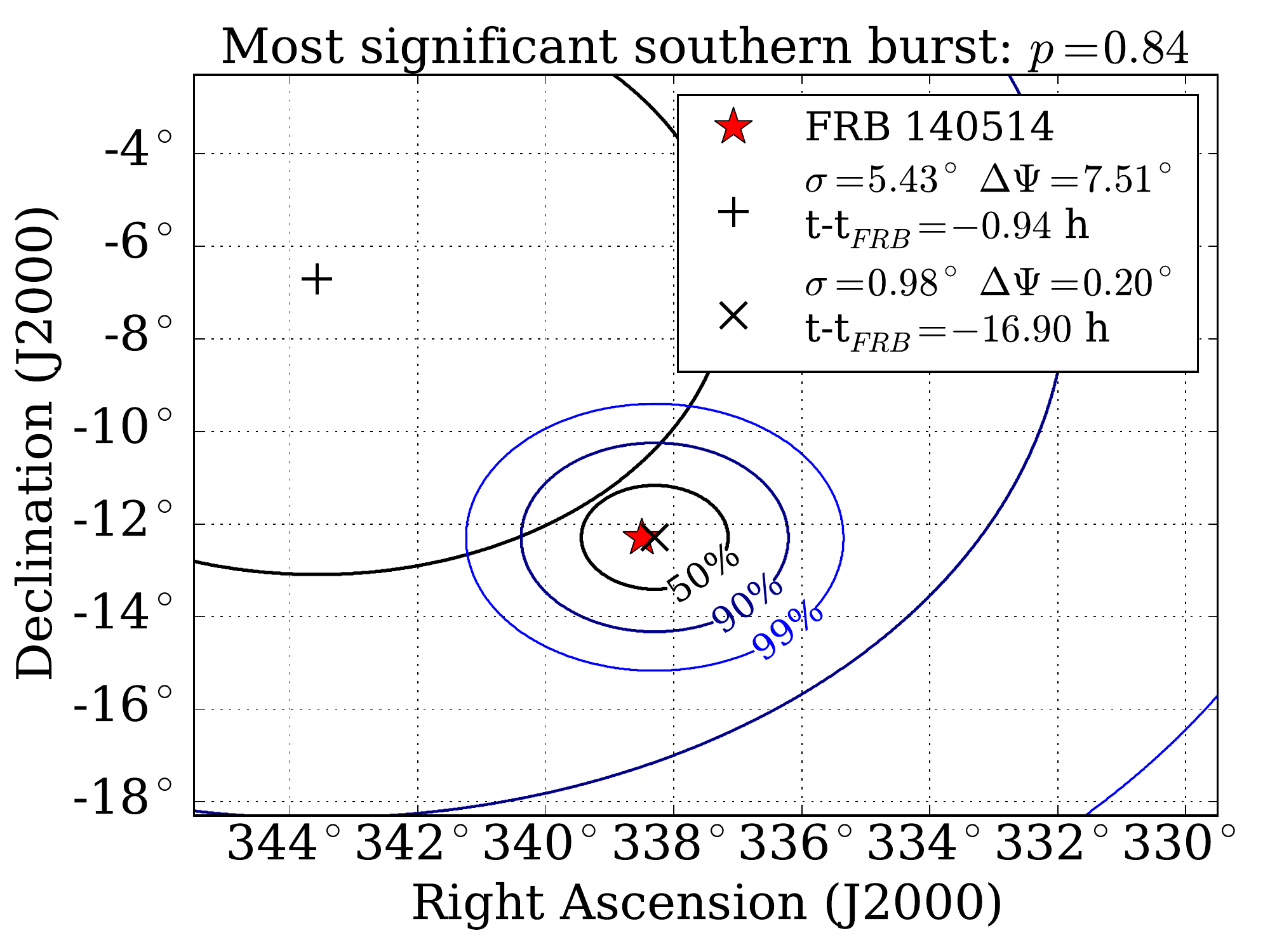}
  \end{minipage}
  \vspace{.1in}
  \caption{\emph{Left}: The most signal-like event in both northern searches was detected 200.806 s after the radio detection of FRB 121102 b3. The directional reconstruction of this event has an angular separation $\Delta \Psi = 2.31^\circ$ with the FRB and an estimated error $\sigma = 1.31^\circ$. Event reconstruction contours are drawn for confidence intervals of 50\%, 90\%, and 99\%, taking the reconstruction as a radially symmetric 2-D Gaussian. FRB directional uncertainty ($\ll 1^{\circ}$) is taken into account in this analysis, but not shown for this scale. The post-trial p-value for this max-burst search is $p = 0.25$. 
\emph{Right}: The most signal-like event in the southern searches was coincident with FRB 140514, with which two events' 90\%-confidence intervals overlap. One event was detected 0.94 hours before the detection of FRB 140514 with reconstructed angular separation $\Delta \Psi = 7.51^\circ$ and estimated error $\sigma = 5.43^\circ$. The second was detected only in the largest time window, 16.90 hours before the FRB, with $\Delta \Psi = 0.20^\circ$ and $\sigma = 0.98^\circ$. Although this event appears remarkably coincident with the location of FRB 140514, its significance suffers from the high background rate of the time window in which it first appears. Its angular uncertainty is also roughly twice the median angular uncertainty of its background sample, reducing the contribution its signal PDF $S(x_i)$ has on the TS value. The post-trial p-value for this max-burst search is $p = 0.84$.}
  \label{skymaps}
\end{figure}

In the southern hemisphere, the max-burst search returns the most significant pre-trial p-value ($p=0.412$) at ${\Delta\textrm{T}=167772.16~\textrm{s}}$ with TS and $n_s$ of 0.64 and 0.78, for a post-trial p-value of $p=0.84$. In the southern stacking search, no TS value greater than zero was ever obtained for all $\Delta$T. Even for the largest $\Delta$T, where the southern max-burst search returned a positive TS value at one FRB, the order-of-magnitude increase in background for 9 FRBs stacked sufficiently diminished the significance of the events. Analysis results are summarized in Table~\ref{tab:results}, and sky maps of the events which most contributed to the results of each hemisphere are shown in Figure~\ref{skymaps}.

\begin{figure*}
\centering
  \begin{minipage}[b]{0.46\textwidth}
  \centering
  \includegraphics[width=\textwidth]{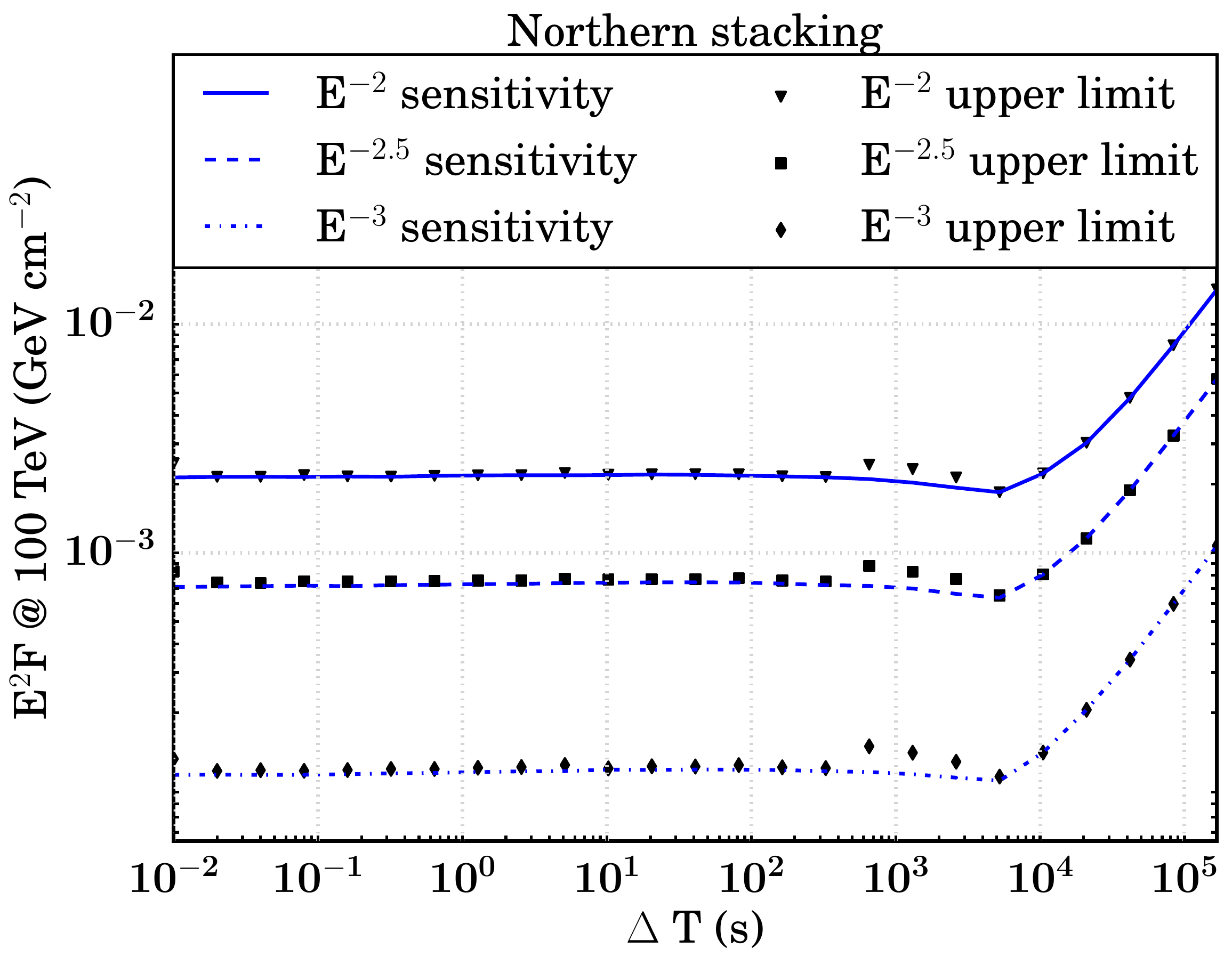}
  \label{fig:limits_N_stacking}
  \end{minipage}
  \quad
  \begin{minipage}[b]{0.46\textwidth}  
  \centering 
  \includegraphics[width=\textwidth]{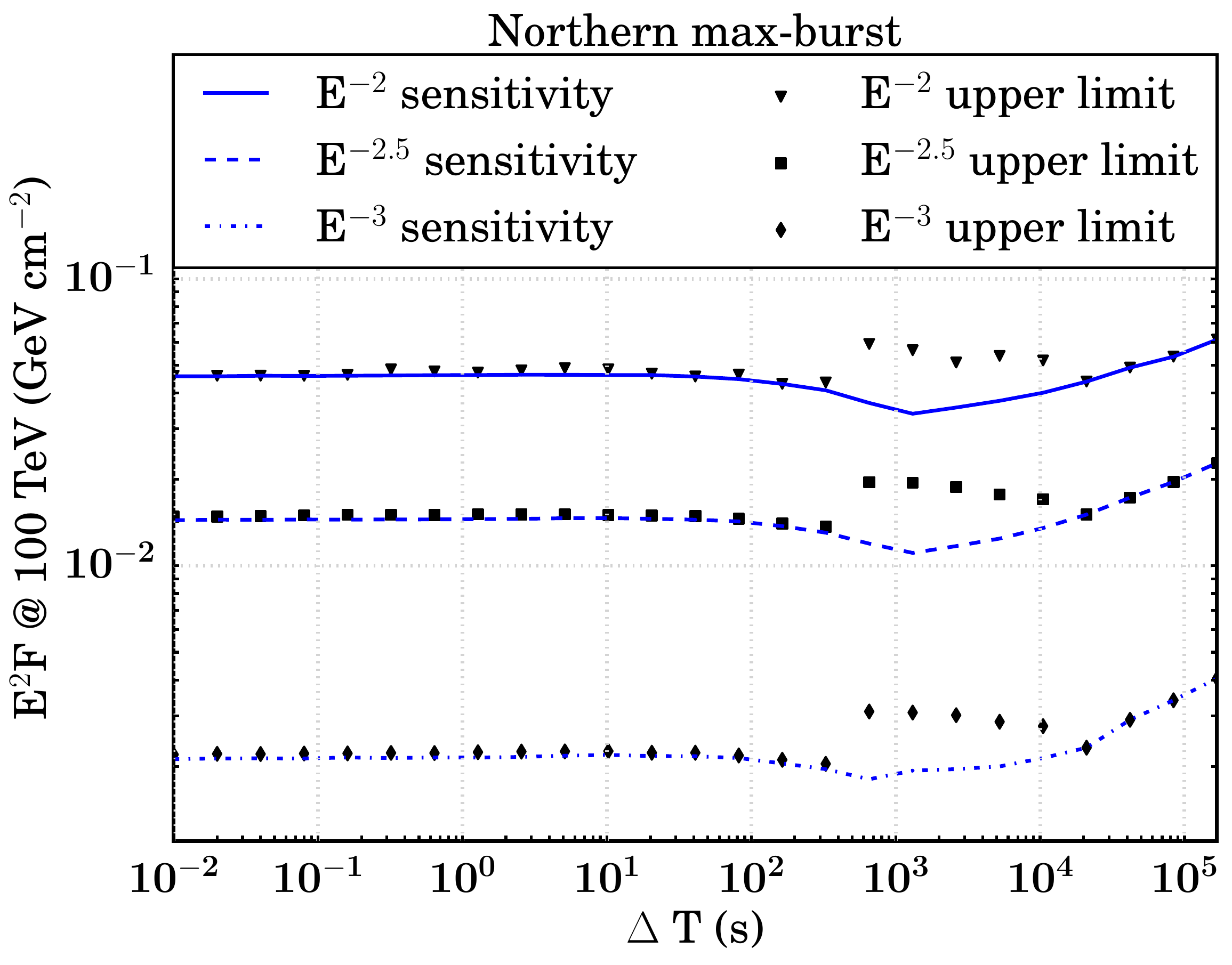} 
  \label{fig:limits_N_maxBurst}
  \end{minipage}
  \quad
  \begin{minipage}[b]{0.46\textwidth}   
  \centering 
  \includegraphics[width=\textwidth]{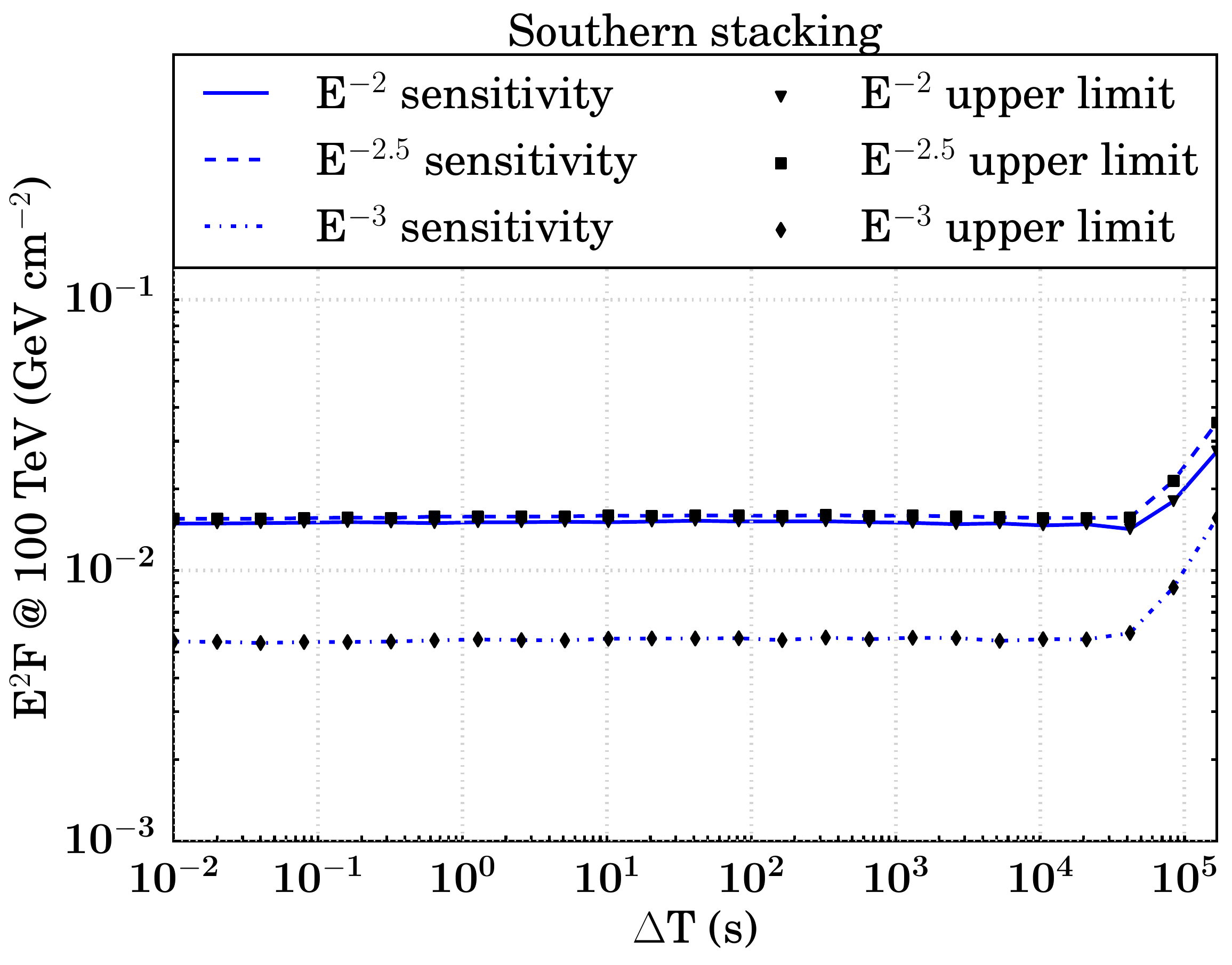}
  \label{fig:limits_S_stacking}
  \end{minipage}
  \quad
  \begin{minipage}[b]{0.46\textwidth}   
  \centering 
  \includegraphics[width=\textwidth]{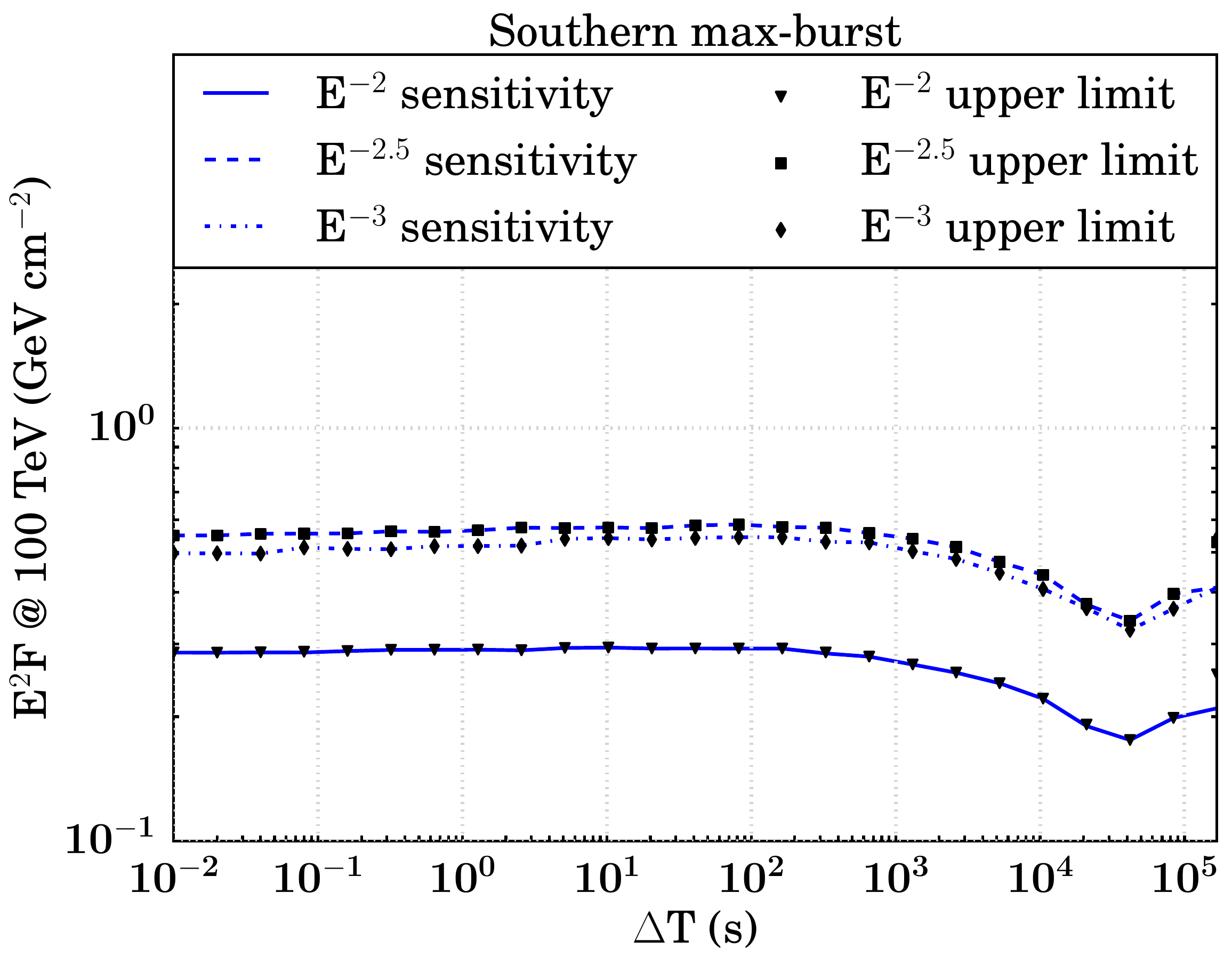}
  \label{fig:limits_S_maxBurst}
  \end{minipage}
  \caption{Sensitivity and upper limits (90\% confidence level) per burst versus $\Delta$T for the stacking and max-burst search in each hemisphere. For the largest $\Delta$T's, in the case that an upper limit fluctuates below the sensitivity, we make the conservative choice to raise the upper limit to the sensitivity value.}
  \label{fig:upperlimits}
\end{figure*}

To set upper limits on the neutrino emission from FRBs, we use the same method which determines sensitivity, using the observed TS rather than the background-only median as a significance threshold. For most $\Delta$T, both the background median and analysis result TS values are zero, resulting in an upper limit equal to the sensitivity (Figure~\ref{fig:upperlimits}). The northern stacking search returned the most constraining 90\% confidence level upper limit for $E^{-2}$ neutrino emission from FRBs among all four searches in this analysis, ${E^2F=0.0021~\textrm{GeV}~\textrm{cm}^{-2}}$ per burst.

This process has been repeated for each source separately to calculate per-burst upper limits (see Table~\ref{tab:perburst_limits}). $E^{-2}$ fluence upper limits were determined by running background and signal-injection trials for a source list containing only one FRB, repeated for each unique source and for each year in which FRB 121102 was detected.

\begin{table}
\caption{Analysis results are summarized for searches in both the northern and southern hemispheres. Where a most significant TS is found, the timing and directional separation of the event which most contributed to that TS value are provided. In the southern stacking test, the TS values for all time windows are zero; there is no $\Delta$T searched which is more signal-like than background-like.}
\vspace{.1in}
\centering
\begin{tabular}{ c c c c c c c c}
\hline
\hline
Northern ($\delta>-5^\circ$) & best fit TS & best fit $n_s$ & \thead{most significant event \\ ($t-t_{\mathrm{FRB}}$, $\Delta \Psi$)} & \thead{pre-trial $p$ \\ (post-trial $p$)} & optimal $\Delta$T & coincident FRB \\
\hline
\hline
max-burst test & 3.90 & 0.99 & (+200.806 s, 2.31$^{\circ}$) & \thead{0.034 \\(0.25)} & 655.36 s & \thead{FRB121102 repeater \\2015/06/02 16:38:07.575 UTC} \\
\hline 
stacking test& 1.41 & 1.01 & (+200.806 s, 2.31$^{\circ}$) & \thead{0.074 \\(0.375)}& 655.36 s & \thead{FRB121102 repeater \\2015/06/02 16:38:07.575 UTC}\\
\hline 
\hline
Southern ($\delta<-5^\circ$)& best fit TS & best fit $n_s$ & \thead{most significant event \\ ($t-t_{\mathrm{FRB}}$, $\Delta \Psi$)} & \thead{pre-trial $p$ \\ (post-trial $p$)}& optimal $\Delta$T & coincident FRB \\
\hline
\hline
max-burst test & 0.64 & 0.78 & (-16.9 hrs, 0.20$^{\circ}$) & \thead{0.412 \\ (0.84)}  & 167772.16 s & \thead{FRB 140514 \\
2014/05/14 17:14:11.06 UTC} \\
\hline 
stacking test & 0 & 0 & -- & \thead{1.0 \\ (1.0)} & -- & -- \\
\hline
\end{tabular}
\label{tab:results}
\vspace{.1in}
\end{table}

\begin{table}
\caption{Neutrino fluence upper limits (90\% confidence) are constructed assuming an $E^{-2}$ spectrum.  The limits have been calculated for each burst individually for the $\Delta $T$~=~0.01$~s time window and are shown here as $E^{2}F$. Each burst from FRB 121102 has a limit corresponding to the year of data during which it was detected.}
\vspace{.1in}
\centering
\begin{tabular}{ c c c}
\hline
\hline
FRB & Dec & E$^{-2}$ fluence upper limit (GeV cm$^{-2}$)\\
\hline
\hline
FRB 121002 & -85$^\circ$ 11$^\prime$ & 1.16\\
\hline
FRB 131104 & -51$^\circ$ 17$^\prime$ & 1.03\\
\hline
FRB 110627 & -44$^\circ$ 44$^\prime$ & 0.963\\
\hline
FRB 150418 & -19$^\circ$ 00$^\prime$ & 0.331\\
\hline
FRB 120127 & -18$^\circ$ 25$^\prime$ & 0.318\\
\hline
FRB 110220 & -12$^\circ$ 24$^\prime$ & 0.184 \\
\hline
FRB 140514 & -12$^\circ$ 18$^\prime$ & 0.192\\
\hline
FRB 130626 & -07$^\circ$ 27$^\prime$ & 0.153\\
\hline
FRB 130729 & -05$^\circ$ 59$^\prime$ & 0.136\\
\hline
FRB 110703 & -02$^\circ$ 52$^\prime$ & 0.0575\\
\hline
FRB 110523 & -00$^\circ$ 12$^\prime$ & 0.0578\\
\hline
FRB 130628 & 03$^\circ$ 26$^\prime$ & 0.0643\\
\hline
FRB 121102 b0 & 33$^\circ$ 05$^\prime$ & 0.0932\\
\hline
FRB 121102 b1-b2 & 33$^\circ$ 05$^\prime$ & 0.0925\\
\hline
FRB 121102 b3-b16 & 33$^\circ$ 05$^\prime$ & 0.0919\\
\hline
LOFAR transient & 86$^\circ$ 22$^\prime$ & 0.164\\
\hline
\end{tabular}
\label{tab:perburst_limits}
\vspace{.1in}
\end{table}

\begin{figure*}[b]
\centering
  \begin{minipage}[b]{0.48\textwidth}
  \centering
  \includegraphics[width=\textwidth]{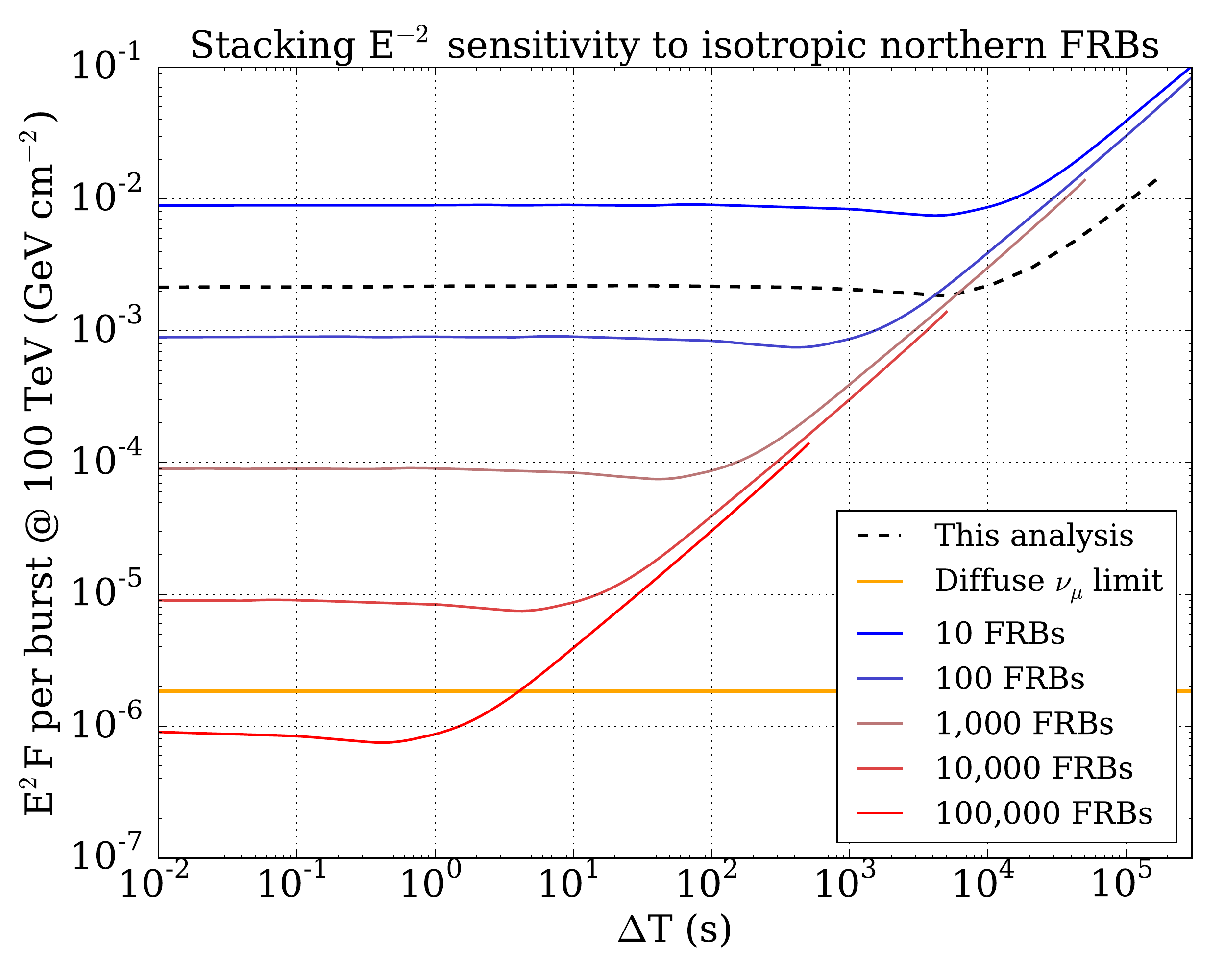}
  \label{fig:limits_Nbursts_N_stacking}
  \end{minipage}
  \quad
  \begin{minipage}[b]{0.48\textwidth}  
  \centering 
  \includegraphics[width=\textwidth]{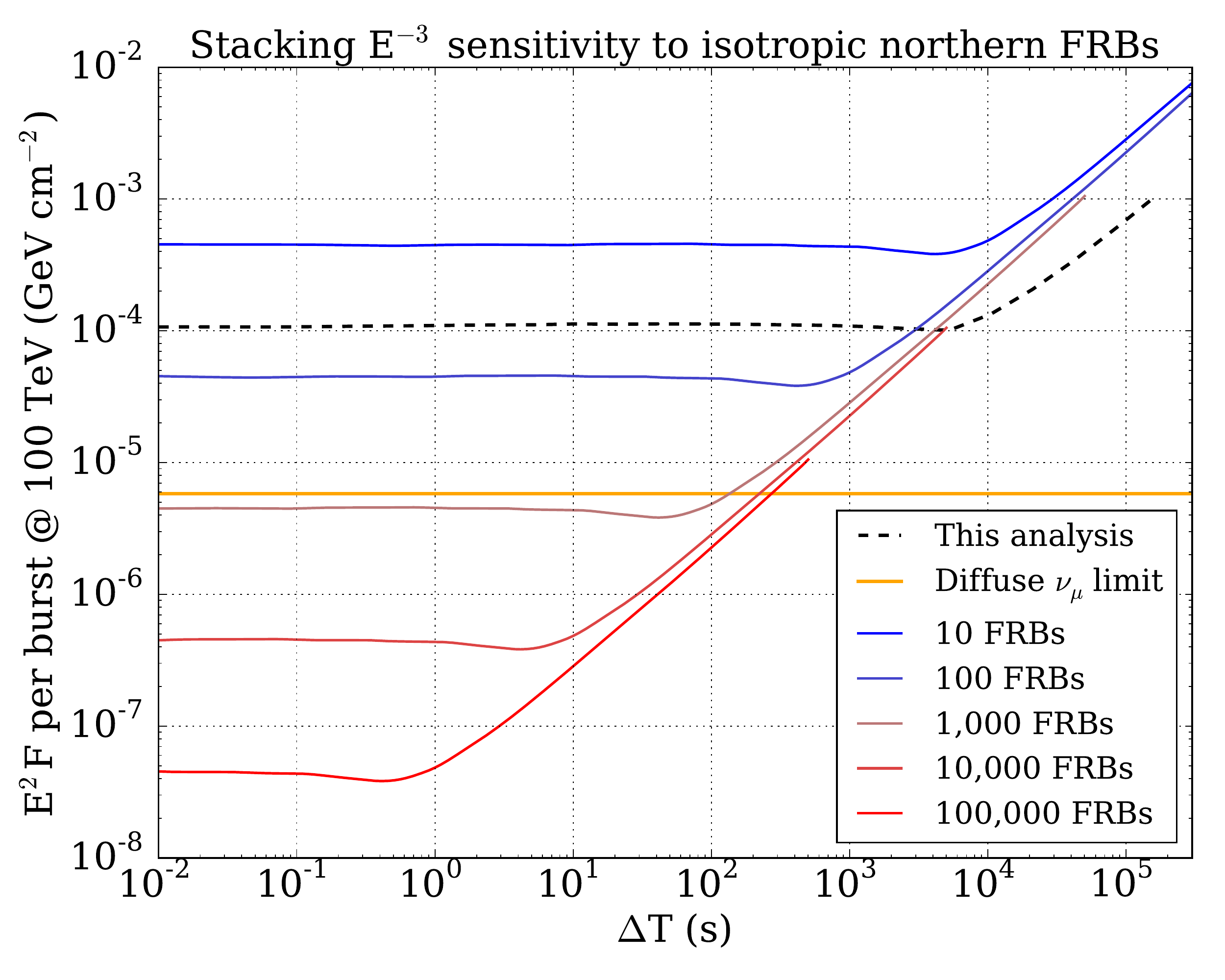}
  \label{fig:limits_Nbursts_N_maxBurst}
  \end{minipage}
  \quad
  \begin{minipage}[b]{0.48\textwidth}   
  \centering 
  \includegraphics[width=\textwidth]{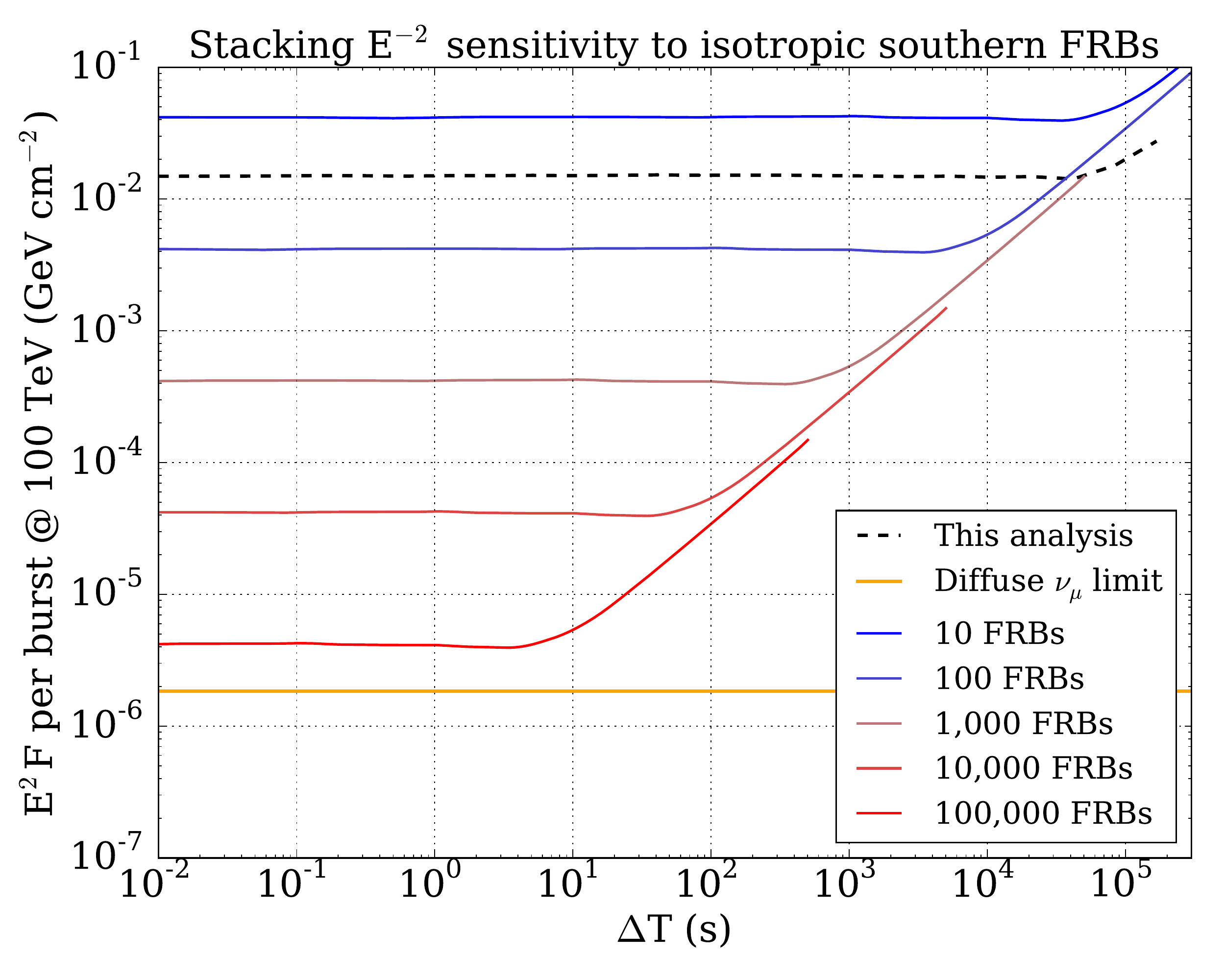}
  \label{fig:limits_Nbursts_S_stacking}
  \end{minipage}
  \quad
  \begin{minipage}[b]{0.48\textwidth}   
  \centering 
  \includegraphics[width=\textwidth]{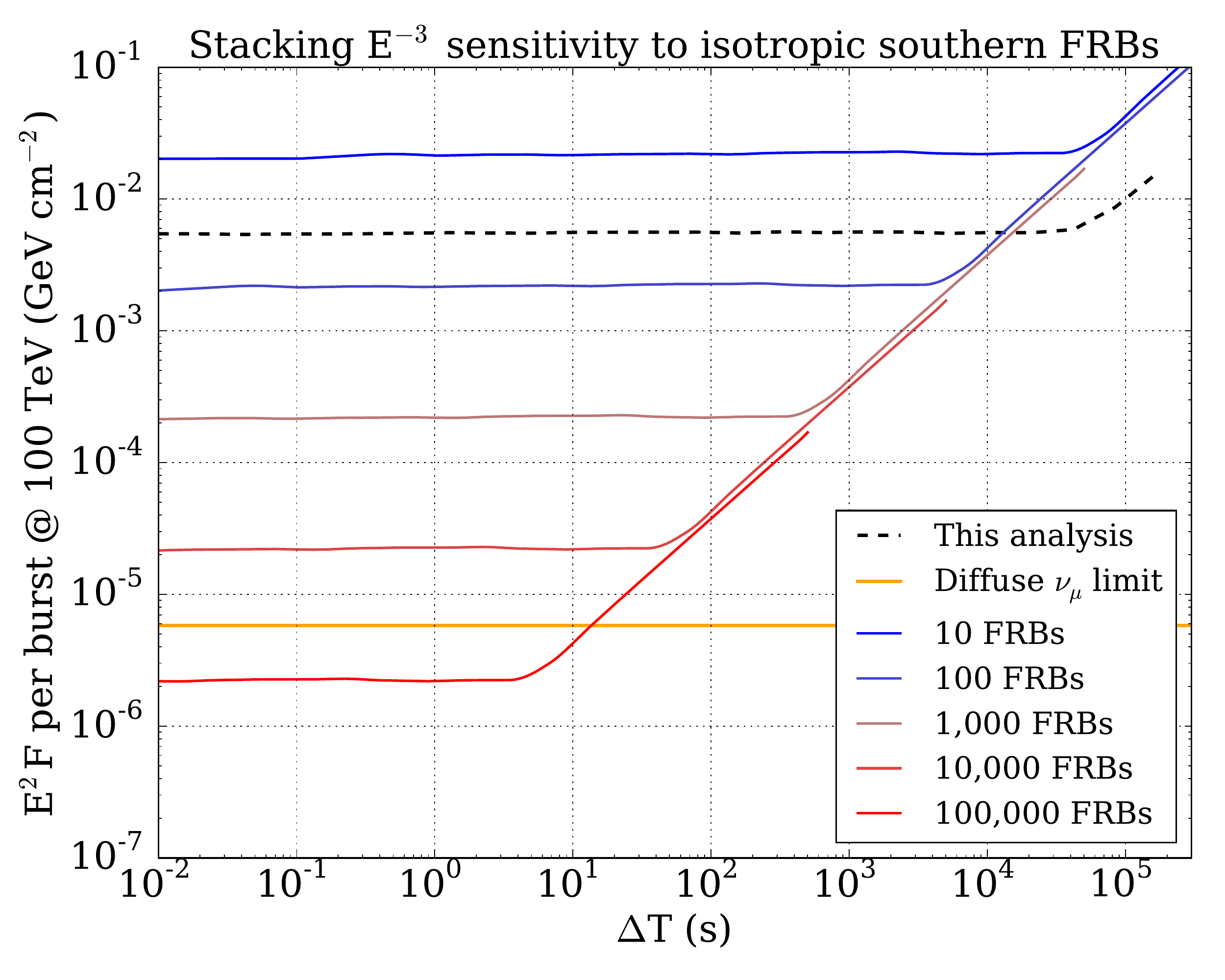}
  \label{fig:limits_Nbursts_S_maxBurst}
  \end{minipage}
  \caption{The stacking sensitivity to FRBs relies on the number and locations of sources detected. Since the list of detected FRBs is expected to grow exponentially in the coming years and without significant directional bias, the per-burst sensitivity to an isotropic hemisphere of FRBs has been calculated for a range of source list sizes. Sensitivity vs. $\Delta$T is shown for two emission spectra, $E^{-2}$ and $E^{-3}$, in each hemisphere for source list sizes ranging from $10-100,000$ FRBs. The respective stacking sensitivities from this analysis are overlaid for comparison, with total fluence divided by the number of sources -- 9 in the south, 20 in the north -- for \emph{per-burst} fluence. These sensitivities outperform the expected sensitivity to an isotropic sky because the FRBs in this analysis were of higher-than-isotropic declination on average. Since our background rates peak at the horizon, the rate of coincident background events in stacking trials was lower than would be expected from an isotropic distribution of FRBs as well. This lowers the baseline for the stacking sensitivity curve and moves the up-turn at large $\Delta$T to the right, as shown by the crossover near $10^4$ s in each plot. For comparison, the limits set by constraining the total all-sky FRB fluence to be less than or equal to IceCube's astrophysical $\nu_{\mu}$ flux are provided, assuming an FRB occurrence rate of 3,000 sky$^{-1}$ day$^{-1}$. With data optimized specifically for sensitivity to FRBs and an orders-of-magnitude larger FRB source list, we expect future limits to improve upon those set by IceCube's diffuse astrophysical neutrino flux.}
  \label{sensitivity_Nbursts}
\end{figure*}

\section{Conclusion and Outlook} \label{conclusion}
In a search for muon neutrinos from 29 FRBs detected from 2010 May 31 to 2016 May 12, no significant correlation has been found. In both hemispheres, several events were found to be spatially coincident with some FRBs but also consistent with background.

Therefore, we set upper limits on neutrino emission from FRBs as a function of time window searched. For a $E^{-2}$ energy spectrum, the most stringent limit on neutrino fluence per burst is ${E^2F = 0.0021~\textrm{GeV}~\textrm{cm}^{-2}}$, obtained from the shortest time window (10 ms) in the northern stacking search. This limit is much improved in comparison to a previous search with only one year of IceCube data and using a binned likelihood method \citep{Fahey:2016czk}. The limits set in this paper are also the most constraining ones on neutrinos from FRBs for neutrino energies above 1 TeV.

At the moment, we can set even more constraining limits on high-energy neutrino emission from FRBs using IceCube's astrophysical $\nu_{\mu}$ flux measurement~\citep{numu6year}, assuming the current catalog of detected FRBs is representative of a homogeneous source class. Using an estimated all-sky FRB occurrence rate of 3,000 sky$^{-1}$ day$^{-1}$~\citep{Macquart:2017tok}, the $\nu_{\mu}$ fluence per FRB at 100 TeV cannot exceed $E^2F = 1.9\cdot10^{-6}$~GeV~cm$^{-2}$ for an emission spectrum of $E^{-2}$; otherwise, FRBs would contribute more than the entire measured astrophysical $\nu_{\mu}$ flux. The astrophysical flux used here is extrapolated from a fit at energies of 194 TeV -- 7.8 PeV, so it is only a rough estimate of the maximum neutrino emission from FRBs in the energy range this analysis concerns.

\begin{figure}[ht!]
\centering
\begin{minipage}[b]{0.49\textwidth}
    \includegraphics[width=\textwidth]{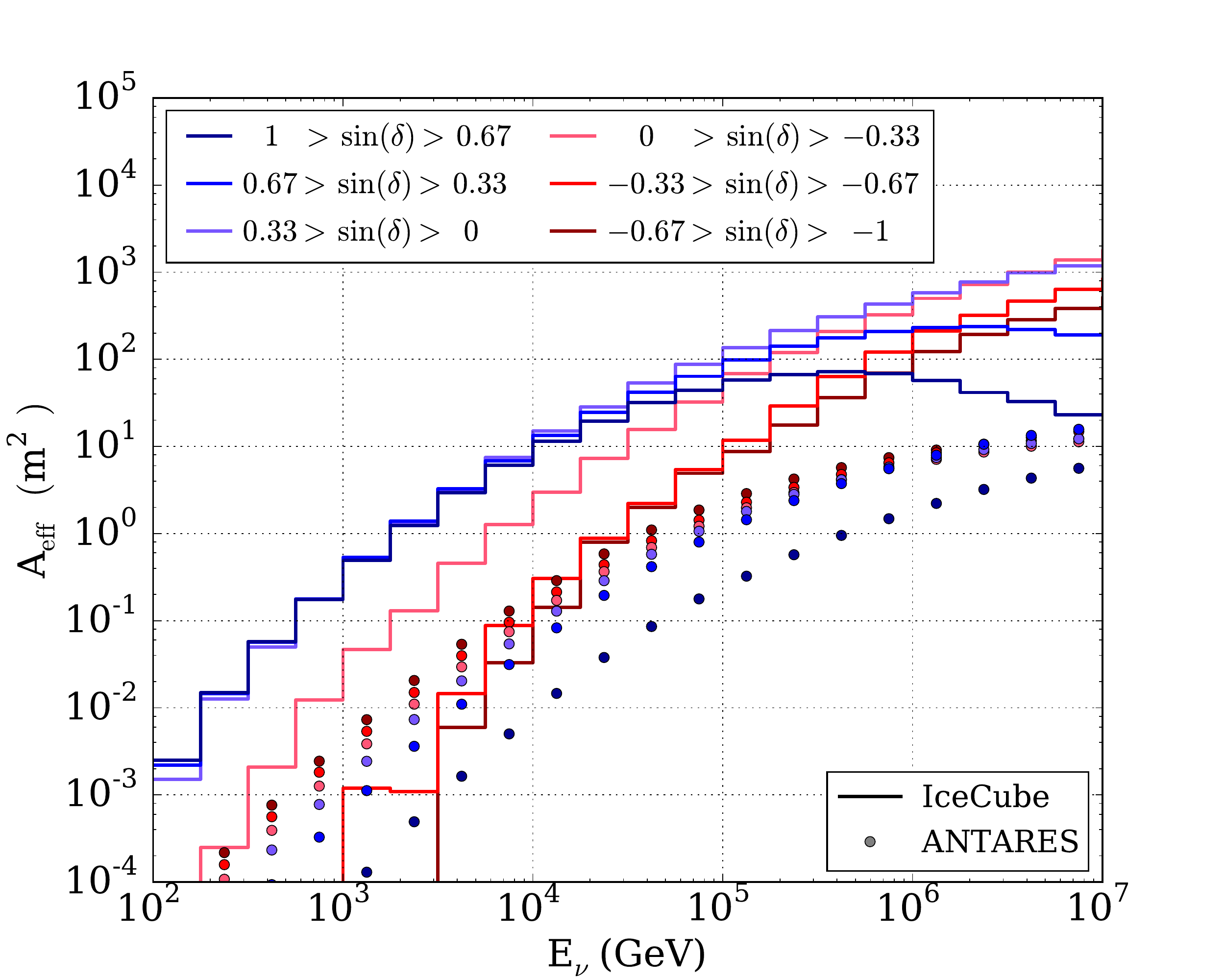}
  \end{minipage}
  \hfill
  \begin{minipage}[b]{0.49\textwidth}
    \includegraphics[width=\textwidth]{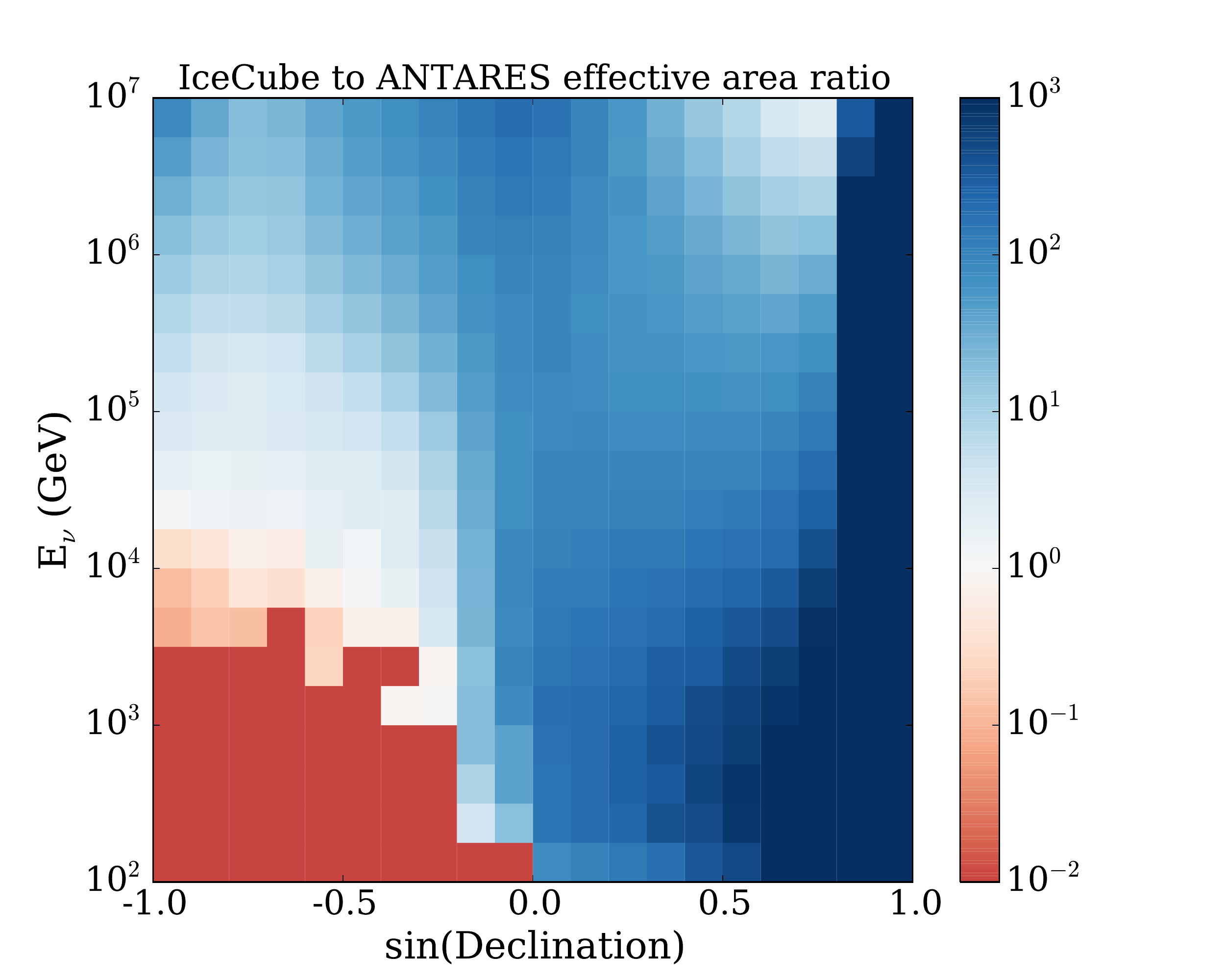}
  \end{minipage}
  \caption{\emph{Left}: The effective area of IceCube to muon neutrinos with energies 100 GeV - 10 PeV is shown for the event selection applied to this analysis' data samples. For comparison, the effective area of the ANTARES observatory's point-source event selection is shown over the same range (circles). Below 1 TeV, the effective area of ANTARES is greater for most of the southern sky and that of IceCube dominates in the north. Above 50 TeV, IceCube's effective area dominates in all declinations in the energy range for which data are available. \emph{Right}: A 2-dimensional plot shows the ratio of the effective areas of IceCube to ANTARES over energy and declination, with a bin-width of 0.1 in $\sin(\delta)$ and bin-height equal to one quarter of a decade in energy. Where ANTARES provides a non-zero effective area, but IceCube's is equal to zero for this event selection, the ratio plotted is the scale minimum $10^{-2}$; likewise, where the converse is true, the ratio plotted is the scale maximum $10^{3}$.}
  \label{Aeff_ANTARES_sidebyside}
\end{figure}

With newly operating radio observatories like CHIME~\citep{0004-637X-844-2-161}, we expect on the order of 1,000 FRBs to be discovered quasi-isotropically each year, which will improve the sensitivity of IceCube to a follow-up stacking search by orders of magnitude (Figure~\ref{sensitivity_Nbursts}). Future analyses using IceCube data may also benefit from a more inclusive dataset, allowing a higher overall rate of muon-like and cascade-like events in exchange for increased sensitivity at $\Delta$T$~<~1,000$ s. Cascade-like events do not contain muons, and as a result provide an angular resolution on the order of $10^\circ$. However, a coincident event may still provide potential for high significance in very short time windows, where background is low. Furthermore, if some sub-class of FRBs is associated with nearby supernovae, MeV-scale neutrinos can be searched in the IceCube supernova stream which looks for a sudden increase in the overall noise rate of the detector modules~\citep{Abbasi:2011ss}. 

The ANTARES neutrino observatory is most sensitive in the southern hemisphere, where the majority of FRB sources have been detected to date. Higher FRB detection rate (due to more observation time) from the southern hemisphere also provides ANTARES the opportunities for rapid follow-up observations when FRBs are caught in real time \citep{Petroff:2017pff}. However, we emphasize that IceCube also has excellent sensitivity in much of the southern hemisphere. In Figure~\ref{Aeff_ANTARES_sidebyside}, we provide a quantitative comparison of the effective areas of the two observatories, which can serve as a useful reference when future FRBs are detected at arbitrary declinations. At energies above 50 TeV, the effective area of IceCube to neutrinos is the highest of any neutrino observatory across the entire ($4\pi$) sky (Figure~\ref{Aeff_ANTARES_sidebyside}). For ${E_{\nu}<50}$ TeV, particularly where ${\sin(\delta)<-0.33}$, ANTARES complements IceCube in searches for isotropic transient sources, achieving greater effective area in $1/3$ of the sky. Since ANTARES is not located at a pole, the zenith angle of any astrophysical source changes throughout the day, thus detector overburden and sensitivity are time-dependent. Therefore, the effective areas provided by ANTARES for a given declination band are the day-averaged values \citep{Adrian-Martinez:2014wzf}. A joint stacking analysis between IceCube and ANTARES \citep{Adrian-Martinez:2016xgn, Adrian-Martinez:2015ver} could maximize the sensitivity of neutrino searches from FRBs across the full sky. Furthermore, with the implementation of the expanding time window techniques, IceCube can now follow up on generic fast transients rapidly, enabling monitoring of the transient sky in the neutrino sector \citep{Aartsen:2016lmt}.

\acknowledgments
The authors gratefully acknowledge the support from the following agencies and institutions: USA – U.S. National Science Foundation-Office of Polar Programs, U.S. National Science Foundation-Physics Division, Wisconsin Alumni Research Foundation, Center for High Throughput Computing (CHTC) at the University of Wisconsin–Madison, Open Science Grid (OSG), Extreme Science and Engineering Discovery Environment (XSEDE), U.S. Department of Energy–National Energy Research Scientific Computing Center, Particle astrophysics research computing center at the University of Maryland, Institute for Cyber-Enabled Research at Michigan State University, and Astroparticle physics computational facility at Marquette University; Belgium – Funds for Scientific Research (FRS-FNRS and FWO), FWO Odysseus and Big Science programmes, and Belgian Federal Science Policy Office (Belspo); Germany – Bundesministerium für Bildung und Forschung (BMBF), Deutsche Forschungsgemeinschaft (DFG) and the German Excellence Initiative, Helmholtz Alliance for Astroparticle Physics (HAP), Initiative and Networking Fund of the Helmholtz Association, Deutsches Elektronen Synchrotron (DESY), and High Performance Computing cluster of the RWTH Aachen; Sweden – Swedish Research Council, Swedish Polar Research Secretariat, Swedish National Infrastructure for Computing (SNIC), and Knut and Alice Wallenberg Foundation; Australia – Australian Research Council; Canada – Natural Sciences and Engineering Research Council of Canada, Calcul Québec, Compute Ontario, Canada Foundation for Innovation, WestGrid, and Compute Canada; Denmark – Villum Fonden, Danish National Research Foundation (DNRF); New Zealand – Marsden Fund; Japan - Japan Society for Promotion of Science (JSPS) and Institute for Global Prominent Research (IGPR) of Chiba University; Korea – National Research Foundation of Korea (NRF); Switzerland – Swiss National Science Foundation (SNSF).


\bibliographystyle{yahapj}
\bibliography{references}

\end{document}